\documentclass[acmlarge]{acmart}

\usepackage[utf8]{inputenc}
\usepackage[T1]{fontenc}
\usepackage{graphicx}
\usepackage{adjustbox}
\usepackage{tabularx}
\usepackage{subcaption}
\usepackage[english]{babel}
\usepackage[figuresright]{rotating}
\usepackage{makecell}
\usepackage{array}
\usepackage{caption}
\usepackage{multirow}
\usepackage{url}
\usepackage{float}
\usepackage{hyperref}
\usepackage{adjustbox}
\usepackage{threeparttable}
\usepackage{rotating} 
\usepackage{array}
\usepackage{makecell}
\usepackage{tikz}
\usepackage{afterpage}
\usepackage{longtable,lscape}
\usepackage{enumitem}
\usepackage{color, colortbl}
\usepackage{wrapfig}
\usepackage[linesnumbered]{algorithm2e}
\definecolor{Gray}{gray}{0.9}
\definecolor{LightGray}{gray}{0.6}
\usepackage[most]{tcolorbox}
\hypersetup{colorlinks, citecolor={black}, filecolor=black, linkcolor=black, urlcolor=black}

\usepackage{xcolor}
\definecolor{green(munsell)}{rgb}{0.0, 0.66, 0.47}
\definecolor{cadmiumgreen}{rgb}{0.0, 0.42, 0.24}
\definecolor{cobalt}{rgb}{0.0, 0.28, 0.67}
\definecolor{amber(sae/ece)}{rgb}{1.0, 0.49, 0.0}
\newlength\MAX  

\newlength\BARSIZE  

\newcommand*\ChartBarBlue[1]{\textcolor{cobalt}{\rule{\BARSIZE}{2ex}}}
\newcommand*\ChartBarGreen[1]{\textcolor{green(munsell)}{\rule{\BARSIZE}{2ex}}}
\newcommand*\ChartBarOrange[1]{\textcolor{amber(sae/ece)}{\rule{\BARSIZE}{2ex}}}

\AtBeginDocument{%
  \providecommand\BibTeX{{%
    \normalfont B\kern-0.5em{\scshape i\kern-0.25em b}\kern-0.8em\TeX}}}
    
\usepackage{soul}

\definecolor{new}{rgb}{0.0, 0.0, 0.0}
\definecolor{edits}{rgb}{255.0, 0.0, 0.0}

\setcopyright{acmcopyright}
\copyrightyear{2022}
\acmYear{2022}
\acmDOI{00.0000/0000000.0000000}
\acmJournal{HEALTH}
\acmVolume{00}
\acmNumber{0}
\acmArticle{000}
\acmMonth{0}

\setcopyright{none}

\makeatletter
\def\@fnsymbol#1{\ensuremath{\ifcase#1\or \dagger\or 
   \mathsection\or \mathparagraph\or \|\or **\or \dagger\dagger
   \or \ddagger\ddagger \else\@ctrerr\fi}}
\makeatother

\newcommand{\lakmal}[1]{\textcolor{black}{#1}}
\newcommand{\wageesha}[1]{\textcolor{black}{#1}}

\begin{document}

\title{Inferring Mood-While-Eating with Smartphone Sensing and Community-Based Model Personalization}

\author{Wageesha Bangamuarachchi}
\authornote{co-primary, listed alphabetically}
\authornote{now at University of Utah, USA}
\affiliation{
  \institution{University of Moratuwa, Sri Lanka}
}

\author{Anju Chamantha}
\authornotemark[1]
\affiliation{
  \institution{University of Moratuwa, Sri Lanka}
}

\author{Lakmal Meegahapola}
\authornotemark[1]
\authornote{now at ETH Zurich, Switzerland}
\email{lmeegahapola@idiap.ch}
\affiliation{\institution{Idiap Research Institute \& EPFL, Switzerland}}

\author{Haeeun Kim}
\affiliation{\institution{Idiap Research Institute \& EPFL, Switzerland}}

\author{Salvador Ruiz-Correa}
\affiliation{
  \institution{IPICYT, Mexico}
}

\author{Indika Perera}
\affiliation{
  \institution{University of Moratuwa, Sri Lanka}
}

\author{Daniel Gatica-Perez}
\affiliation{\institution{Idiap Research Institute \& EPFL, Switzerland}}


\renewcommand{\shortauthors}{Bangamuarachchi, Chamantha, Meegahapola, et al.}

\begin{abstract}

The interplay between mood and eating episodes has been extensively researched within the fields of nutrition, psychology, and behavioral science, revealing a connection between the two. Previous studies have relied on questionnaires and mobile phone self-reports to investigate the relationship between mood and eating. In more recent work, phone sensor data has been utilized to characterize both eating behavior and mood independently, particularly in the context of mobile food diaries and mobile health applications. However, current literature exhibits several limitations: a lack of investigation into the generalization of mood inference models trained with data from various everyday life situations to specific contexts like eating; an absence of studies using sensor data to explore the intersection of mood and eating; and inadequate examination of model personalization techniques within limited label settings, a common challenge in mood inference (i.e., far fewer negative mood reports compared to positive or neutral reports). In this study, we examined the everyday eating and mood using two separate datasets from two different studies: \textit{i)} Mexico (N$_{MEX}$ = 84, 1843 mood-while-eating reports with a label distribution of positive: 51.7\%, neutral: 38.6\% and negative: 9.8\%) in 2019, and \textit{ii)} eight countries (N$_{MUL}$ = 678, 329K mood reports, including 24K mood-while-eating reports with a label distribution of positive: 83\%, neutral: 14.9\%, and negative: 2.2\%) in 2020, which contain both passive smartphone sensing and self-report data. Our results indicate that generic mood inference models experience a decline in performance in specific contexts, such as during eating, highlighting the issue of sub-context shifts in mobile sensing. Moreover, we discovered that population-level (non-personalized) and hybrid (partially personalized) modeling techniques fall short in the commonly used three-class mood inference task (positive, neutral, negative). Additionally, we found that user-level modeling posed challenges for the majority of participants due to insufficient labels and data in the negative class. To overcome these limitations, we implemented a novel community-based personalization approach, building models with data from a set of users similar to the target user. Our findings demonstrate that mood-while-eating can be inferred with accuracies 63.8\% (with F1-score of 62.5) for the MEX dataset and 88.3\% (with F1-score of 85.7) with the MUL dataset using community-based models, surpassing those achieved with traditional methods.

\end{abstract}



\maketitle

\normalsize
\section{Introduction}\label{sec:introduction}

Mood can be defined as "a conscious state of mind or predominant emotion" \cite{MoodDefinition2022}. Even though it is an internal and subjective state, it often can be inferred from behaviors of individuals \cite{servia2017mobile, likamwa2013moodscope}. Mood affects many facets of our daily lives. While positive moods increase the likelihood of sociability, creativity, and planning \cite{diener2015people}, negative moods could alter such behaviors in the short term and also provide a way to adverse health outcomes in the long term \cite{hockey2000effects, baker1995body}. Due to these reasons, understanding the causes and contexts of mood and assessing mood have been active research topics in mobile sensing during the last decade \cite{Meegahapola2021Survey}. Hence, studies have used audio sensors in the smartphone \cite{rachuri2010emotionsense}, phone usage patterns \cite{likamwa2013moodscope}, and multi-modal sensors \cite{meegahapola2023generalization, servia2017mobile, ma2012daily} for this task. Recent work has also used sensor data from wearables to recognize different moods~\cite{li2020extraction} and other mental well-being related aspects~\cite{kodikara2025fatiguesense}. These studies show promise for building mobile health systems that could provide feedback and interventions by considering user moods more meaningfully. However, a majority of mood-related studies have focused on everyday life behavior, and there is not much knowledge on the location, social context, and concurrent activities (i.e., studying, working, eating, etc.) done while mood reports were captured using self-reports, as studies that inspect such aspects are scarce \cite{morshed2022food}.

Mental well-being and eating behavior have been extensively studied, both as individual phenomena and at their intersection. Research has explored the impact of eating behavior on mood~\cite{christensen1993effects}, the relationship between obesity and mental health~\cite{stunkard1959eating}, and the psychological states and routines associated with eating~\cite{warde2016practice}. This growing body of literature underscores the connection between mood and eating, highlighting how understanding this relationship can contribute to a deeper comprehension of both long-term and short-term eating behaviors and lead to the development of more effective practices, interventions, and treatments~\cite{giannopoulou2020mindfulness}. For example, negative moods are known to have a link to bulimic eating episodes, while positive moods are associated with a reduced likelihood of unhealthy eating episodes~\cite{christensen1993effects}. Negative moods have also been found to predict disordered eating episodes, particularly in populations like college students~\cite{johnson1984mood, abraham1982patients, pyle1981bulimia, greenberg1986predictors}, with one study suggesting that mood shifts are the most accurate predictor of disordered eating in this group~\cite{greenberg1987affective}. In addition, numerous studies have demonstrated the influence of dietary habits on mood. For instance, fasting and low carb diets have been associated with improved mood and even therapeutic benefits for chronic pain management in the long run~\cite{Michalsen2010, Fond2013, westman2002effect}.

Specific behaviors during eating episodes, such as consuming calming foods, have been shown to reduce tension, anger, and confusion, while fostering more positive emotional states~\cite{Hussin2013}. Gut health, which is influenced by dietary patterns, has a direct impact on brain function and cognitive performance~\cite{Moulton2019}. Furthermore, clinical and preclinical research has established that dietary habits play a key role in modulating anxiety, stress, and depression~\cite{Luqman2024}, emphasizing the connection between mental health and eating behavior. Understanding mood during eating episodes is particularly important because general mood and specific emotions shape eating behaviors~\cite{bisogni2007dimensions}. For instance, negativity can lead to the consumption of high-calorie, palatable foods, potentially contributing to obesity and eating disorders~\cite{rangan2015electronic}, whereas positivity is linked to healthier eating patterns~\cite{ha2023role}. Therefore, recognizing mood in the context of eating episodes enables personalized interventions, such as real-time feedback to mitigate emotional eating, thereby enhancing both mental and physical health. The bidirectional relationship between mood and eating behavior underscores the need for context-specific mood recognition to inform effective strategies for promoting overall well-being~\cite{ljubivcic2023emotions}. In this context, accurate mood inferences during eating hold value, particularly for mobile food diaries and digital health applications. Such systems not only track what individuals are eating but also capture their mental state and contextual aspects while eating, providing a holistic view of eating behavior~\cite{bisogni2007dimensions, Biel2018, morshed2022food, meegahapola2021one, Meegahapola2020Alone}.

Recent studies in mental well-being and mobile sensing highlight the challenge of model generalization across diverse contexts, such as different countries and time periods, due to data distribution disparities, often referred to as distribution shifts~\cite{meegahapola2023generalization, xu2023globem, KarimAssi2023}. While much work has focused on generalization across source and target domains with minimal overlap~\cite{nanchen2023keep, chang2020systematic}, the issue of sub-sample and sub-population shifts~\cite{jones2020selective, yang2023change} has been studied in fields like computer vision and natural language processing but remains underexplored in mobile sensing. In particular, there is limited research on generalization performance for sensing-based mood inference models within specific situational contexts, such as eating, studying, or working. Recent efforts emphasize the importance of context-specific modeling, with studies examining workplace stress~\cite{9953824Morshed}, psychological well-being during work~\cite{nepal2022survey}, and mood during activities like driving~\cite{van2013using, van2012influence}, underscoring the value of understanding mood in specific settings. However, the performance of generic mood inference models in specific contexts, such as eating episodes, and the development of context-specific models for such situations remain largely unexplored. Although research on context-agnostic mood inference exists~\cite{likamwa2013moodscope, servia2017mobile}, its application to specific contexts like eating is limited. Further research in this area is essential to develop robust passive mood-tracking systems that generalize effectively across everyday situations, with significant implications for personalized health and well-being interventions.

Smartphone sensing has demonstrated promise in mood-related tasks, such as inferring positive and negative moods~\cite{servia2017mobile}, levels of pleasure and activeness~\cite{likamwa2013moodscope}, mood instability~\cite{morshed2019prediction}, and feelings of displeasure, tiredness, and tension~\cite{ma2012daily}. It has also proven effective in eating-related tasks, including identifying eating episodes~\cite{sen2020annapurna, thomaz2015practical}, differentiating between meals and snacks~\cite{Biel2018}, understanding the social context of eating~\cite{Meegahapola2020Alone}, and detecting overeating~\cite{meegahapola2021one}. However, the intersection of mood and eating has received comparatively little attention, despite its potential to enhance mobile health applications with context-aware features. The ability to infer an individual's mood during eating episodes could significantly improve targeted feedback or interventions. For example, frequent snacking, often linked to unhealthy eating habits, could be addressed with timely interventions during vulnerable periods~\cite{nahum2018just}. The effectiveness of such interventions depends on factors like the individual's mood and the social context of the eating episode that determine the receptivity.

Finally, in the context of machine learning-based modeling, the traditional approach in mobile sensing studies has been to use one-size-fits-all models, which have shown effectiveness in experimental settings~\cite{xu2021leveraging, kao2020user, KarimAssi2023, meegahapola2023generalization}. However, in real-world scenarios, such models often require personalization techniques to address the heterogeneity of user behaviors~\cite{kao2020user, jiang2018towards, Bangamuarachchi2022Sensing, han2024systematic}. Personalization, however, demands a significant amount of data, which may be unavailable during the initial stages of deployment or even later if users fail to provide sufficient self-reported data as ground truth~\cite{likamwa2013moodscope}. This limitation, commonly referred to as the "cold start problem" or "user-held out" challenge, is a well-documented issue in mobile sensing research~\cite{kao2020user, jaques2015predicting}. Consequently, developing personalized mood inference models that are both robust to user heterogeneity and capable of functioning with limited data remains a significant challenge. Considering these constraints, and in line with the terminology provided in Table~\ref{table:keywords_explaination}, we pose the following research questions.

\begin{itemize}[wide, labelwidth=!, labelindent=0pt]
    \item[\textbf{RQ1:}] Given smartphone sensing data from college students, do generic mood inference models (trained with all available data regardless of the activity performed by users while reporting mood) work well for specific contexts such as eating (i.e., mood-while-eating)?
    \item[\textbf{RQ2:}] Can the self-reported mood of college students during eating be inferred using smartphone sensing data with population-level (non personalized), hybrid (partially personalized), and user-level (fully personalized) models? What are the challenges of making such inferences? 
    \item[\textbf{RQ3:}] What measures should be taken to tackle issues such as lack of individual data and cold-start problem in mood-while-eating inference, in building personalized models? \wageesha{Can a community-based model overcome such issues?}
\end{itemize}{}

By addressing the above research questions, this paper provides the following novel contributions: 
\begin{itemize}[wide, labelwidth=!, labelindent=0pt]
    \item[\textbf{Contribution 1:}] We conducted an analysis using a dataset collected from 678 participants in multiple countries. This dataset consists of mobile sensing data of participants' everyday moods and behavior. First, we trained a generic three-class (positive, neutral, negative) mood inference model. We found that this model does not work similarly across all contexts (i.e., walking, resting, eating, working, etc.) and favors certain contexts ahead of others due to many reasons such as the nature of the activities, their prevalence, etc. In fact, for the eating context, the mood inference performance was below the overall performance. In addition, when increasing the representation of mood-while-eating reports in the training set for the generic mood inference model, the performance of the model on mood-while-eating reports on the testing set increased. However, overall performance declined, indicating the difficulty for the model to generalize to mood reports captured during different situated contexts. These results point towards the need for context-specific models for mood-while-eating inference. \lakmal{Moreover, while this study focused on mood during eating episodes, we believe this finding increases awareness of sub-context generalization issues that could pervasively occur in mobile health sensing and longitudinal behavior modeling.} 
    
    \item[\textbf{Contribution 2:}] We conducted an analysis of the previously mentioned smartphone sensing dataset, and in addition, a second smartphone sensing dataset collected from 84 college students in Mexico during a separate study, regarding their everyday eating behavior. Building on prior work in nutrition and behavioral sciences that emphasized the need to study eating behavior and mood in greater depth, we defined and evaluated a mood-while-eating inference task  (Table~\ref{table:keywords_explaination}). This three-class inference task attempts to infer positive, neutral, and negative moods associated with eating episodes. We show that population-level models (non-personalized) do not generalize well to unseen individuals (accuracy of 24.1\% for the Mexico dataset and 82.4\% for the multi-country dataset). Then, we show that hybrid models (partially personalized) provided an accuracy increase of 30.5\% for the Mexico dataset and 1.3\% for the multi-country dataset. We also discuss how training user-level models is difficult in both datasets, with the lack of individual data from a majority of users for the negative class. 
    
    \item[\textbf{Contribution 3:}] We show that personalization is a key component in achieving higher accuracies because subjective aspects such as mood are hard to generalize using one model that fits all. In addition, we propose an approach of \textit{Community-Based Model Personalization}, building on similarities of users to create a unique community for each target user (Section~\ref{subsec:method_rq3}) for personalized inferences. We show that our approach achieves an accuracy of 63.8\% for the Mexico dataset and 85.7\% for the multi-country dataset.
    
\end{itemize}{}

\lakmal{The paper is organized as follows. In Section~\ref{sub:related_work}, we describe the background and related work. Then we describe the dataset and features used in Section~\ref{sec:mobile_app}. In Section~\ref{sec:rq1_methodd_results}, Section~\ref{sec:rq2_methodd_results} and Section~\ref{sec:rq3_methodd_results} we present method and results of RQ1, RQ2 and RQ3, respectively. We conclude the paper with the Discussion in Section~\ref{sec:discussion}, and the Conclusion in Section~\ref{sec:conclusion}.}

\section{Definitions, Background and Related Work}\label{sub:related_work}

\begin{table*}
\centering
\caption{Terminology and description regarding different model types and Mood-while-Eating. Degree of Personalization increases when going from Population-Level to User-Level.}
\label{table:keywords_explaination}
\resizebox{\textwidth}{!}{%
\begin{tabular}{>{\arraybackslash}m{2.6cm} >{\arraybackslash}m{15cm}}

\rowcolor{gray!20}
\textbf{\makecell[l]{Terminology}}  & 
\textbf{Description} 
\\

\arrayrulecolor{Gray}
\hline

\textcolor{new}{Population-Level} &
\textcolor{new}{The set of users present in the training set and the testing set are disjoint. This represents the case where the model was trained on a subset of the population, and a new set of users joined the system and started using its model. These models do not use end-user data in building the model, and are built with the assumption that a single model will generalize well for a large number of individuals. In practice, this is similar to leave-one-user-out/leave-k-users-out strategy.}\\

\arrayrulecolor{Gray}
\hline 

\textcolor{new}{Hybrid}&
\textcolor{new}{The sets of users in the training and testing splits are not disjoint. Part of the data of some users present in the training set is used in the testing set. This represents the case where the model was trained on the population, and the same people whose data were used in training continue to use the model as a system users. Hence, models are partially personalized. Therefore, models use data from both the individual and others in training. These can also be known as partially personalized models. However, depending on the data coming from others and inter-subject variability, the degree to which personalization works could differ. In practice, this is similar to the leave-one-sample-out/leave-k-samples-out strategy.} \\

\arrayrulecolor{Gray}
\hline 

\textcolor{new}{Community-Based} &
\textcolor{new}{This approach builds on top of hybrid models and uses a part of the target user’s data and data from other users who are similar to the target user (a subset of the population used in hybrid models) in training models. These models do not need a substantial amount of data from the target user for personalization because they also rely on users similar to the target user.} \\

\arrayrulecolor{Gray}
\hline 

\textcolor{new}{User-Level} &
\textcolor{new}{These are also known as fully personalized models, which only use target user’s data in both training and testing. These models work well for individuals with low inter-subject variability, hence generalizing well for a specific individual.  However, these models need a considerable amount of data from each individual for personalization} \\

\arrayrulecolor{gray!15}
\hline 

\textcolor{new}{Mood-while-Eating}&
\textcolor{new}{The mood-while-eating corresponds to the instantaneous \emph{valence} \cite{russell1980circumplex} during eating episodes as reported by study participants on a five-point scale (from very positive to very negative), reduced to a three-point scale corresponding to positive, negative, and neutral classes. Our definition follows prior ubicomp work \cite{likamwa2013moodscope, servia2017mobile, meegahapola2023generalization} that used valence for mood inference. The key difference is that we focus on specific eating episodes and related moods (see Section~\ref{subsec:eatingepisode}).} \\

\arrayrulecolor{Gray}
\hline 
\end{tabular}
}
\end{table*} 

\subsection{{Defining Mood-While-Eating}}\label{subsec:eatingepisode}

\wageesha{Previous research has employed various methods for capturing mental wellbeing-related attributes, including mood, using continuous scales \cite{servia2017mobile}, two-point scales \cite{taylor2017personalized}, and seven-point scales \cite{bogomolov2013happiness}, and has applied them in two/three-class inferences and regression tasks. In this study, we used datasets that reported mood on five-point scales, similar to prior work \cite{zhang2021putting, likamwa2013moodscope}. The responses were: (1) in a very negative mood; (2) in a slightly negative mood; (3) in a neutral mood; (4) in a slightly positive mood; and (5) in a very positive mood. For the proposed three-class classification task, responses 1 and 2 were classified as negative, 3 as neutral, and 4 and 5 as positive. As such, throughout this study, we consider these three levels as a representation of self-reported mood-while-eating, in line with previous studies \cite{wampfler2022affective, soleymani2011multimodal, meegahapola2023generalization, kammoun2023understanding}. Moreover, the limitation with the number of data points in the very negative and slightly negative mood classes was taken into consideration while classifying the three classes. Additionally, there is literature that suggests inferring self-reported mood \cite{suhara2017deepmood} and normalized mood, i.e., since users report their mood differently, users have individual mood distributions, and the user is said to be in a positive mood if a user reports a value higher than the median value of the past distribution \cite{servia2017mobile}. While both of these approaches have distinct advantages, in this paper, we infer self-reported mood without normalization, as the use of a Likert scale makes it difficult to normalize based on individual users or populations. Further, as previous studies have demonstrated \cite{deubler2020k, schouteten2019comparing}, valence may explain a greater portion of a user's emotional response than arousal. For example, as highlighted in \cite{schouteten2019comparing}, the authors attempted to predict the food preferences of children correlated with arousal and valence, gathered using an emoji-based data collection method, and were able to show a high correlation between food selection and emojis, which describe valence levels (positive, negative, and neutral mood). However, even though we acknowledge that both valence and arousal could affect eating behavior, in this study, we only consider valence due to the unavailability of arousal labels in the datasets. Considering all these aspects, \textit{the mood-while-eating corresponds to the instantaneous valence during eating episodes as reported by study participants, converted to a three-point scale corresponding to positive, negative, and neutral classes.}}

\subsection{Theoretical Underpinning for Studying Short-Term Eating Episodes and Mood}

The relationship between eating behavior and mood is complex and driven by various theoretical frameworks and empirical findings. \textit{Emotional eating theory} highlights the link between negative emotions and the consumption of energy-dense, palatable foods, often as a coping strategy during stress~\cite{gibson2006emotional, ljubivcic2023emotions}. This behavior is mediated by the brain's reward circuitry, where emotional regulation and cognitive control intersect with the consumption of palatable foods during eating episodes~\cite{ljubivcic2023emotions, ma2012daily}.

\textit{Physiological mechanisms} also contribute significantly to the interplay between eating and mood. Hormonal regulation, particularly involving ghrelin and leptin in the hypothalamus, plays a critical role in mood and appetite regulation. Dysregulation in these pathways is often linked to mood disorders and unhealthy eating patterns~\cite{ljubivcic2023emotions}. Additionally, the limbic system, which governs emotional processing, promotes motivated behaviors such as eating to alleviate negative emotional states, frequently overriding cognitive control~\cite{godet2022interactions, jurek2024role, stogios2020exploring, bozkurt2024factors}.

From a cognitive perspective, theories such as \textit{restraint theory} and the \textit{affect regulation model} provide further insights. Restraint theory posits that individuals adhering to strict dietary rules are more likely to overeat when experiencing negative emotions, as emotional stress disrupts cognitive control~\cite{reichenberger2020emotional}. The affect regulation model explains how operant and classical conditioning reinforce eating behaviors. Operant conditioning refers to how individuals learn to associate eating with emotional change, such as consuming food as a reward, which reinforces the behavior by reducing negative emotions. Classical conditioning refers to how negative emotions themselves can trigger emotional eating over time. The rewarding aspects of palatable foods alleviate negative emotions (negative reinforcement), while negative emotional states become triggers for eating over time (classical conditioning)~\cite{haedt2011revisiting, reichenberger2020emotional, bozkurt2024factors}.

Individual differences play a significant role in emotional eating behaviors.  Some studies indicate that women are more likely to eat in response to negative emotions, whereas men often consume more during positive emotional states~\cite{zameer2024influence}. Personality traits, such as neuroticism and extroversion, further influence these behaviors, with neurotic individuals showing higher reactivity to stress and extroverts being more prone to social eating~\cite{tinmazouglu2020relationship}.

The bidirectional relationship between mood and eating underscores the importance of this interaction. Unhealthy eating patterns starting at an eating episode level, such as overeating or consuming high-calorie foods, can exacerbate mood disorders like depression, which in turn reproduce unhealthy eating habits~\cite{ljubivcic2023emotions}. Emotional eating and depression are also linked to hedonic eating behaviors, contributing to higher body mass index (BMI) and associated health risks in the long run~\cite{rangan2015electronic, ha2023role}. These findings emphasize the need for comprehensive strategies to address emotional eating at episode level, including interventions targeting reward systems, alternative coping mechanisms, and personalized approaches based on individual differences. Understanding short-term mood dynamics during eating episodes provides valuable insights into long-term eating behavior, with theoretical support from multiple perspectives, as explained above and further explained in Section~\ref{subsec:eating_as_a_holistic}.

\subsection{Mood and Food Consumption}

A substantial body of research has established a strong link between emotional states and eating behaviors~\cite{koster2015mood}. Negative moods or stress often lead to increased consumption of unhealthy, high-fat, and high-sugar foods, while reducing the likelihood of choosing healthier options like fresh produce~\cite{bloemer2018influence, bennett2012perception, wansink2003comfort}. Conversely, positive moods have been associated with increased caloric intake in some contexts, as observed in both regression analyses and controlled experiments~\cite{bongers2013happy, evers2013positive}. Further, mediators of emotional eating have also been explored extensively. For example, high emotional eaters often seek the comforting effects of palatable foods~\cite{strien2019mediation}. Social context is another significant factor, with social settings frequently influencing emotional eating behaviors~\cite{altheimer2019social, raspopow2013social}. The bidirectional relationship between mental health and eating behavior has been extensively studied, with eating patterns often serving as indicators of mental health. For instance, mood changes in bulimic patients and the correlation between binge eating, depression, and stress have been well-documented~\cite{johnson1984mood, pyle1981bulimia, abraham1982patients, greenberg1986predictors, greenberg1987affective}. While these studies have advanced the understanding of the association between mood and eating, most rely on classical methods such as laboratory experiments, mood charting, or studies with clinically diagnosed target groups. These approaches often lack the ecological validity needed to examine everyday associations between food and mood. Recent efforts have sought to bridge this gap using online platforms~\cite{chan2015clickdiary} and mobile sensing technologies~\cite{ashurst2018ema}. However, there remains a research gap in exploring the dynamic relationship between mood and eating behavior using mobile sensing, highlighting the need for further investigation in this domain.

\subsection{Mood and Smartphone Technologies}
Prior studies in mental health have used mobile technologies ranging from text messaging \cite{aguilera2015text} to multi-modal sensors \cite{chan2018asynchronous} for mood tracking. In the early days, the use of mobile apps as a mood chart for clinical purposes was addressed \cite{mark2008chart}. Regarding users of mood-tracking apps, Schueller et al. \cite{schueller2021interview} mentioned that users were primarily motivated by negative events or moods that prompted them to engage in tracking moods to improve the situation. Aligned with their work, we can find a plethora of research that proposed mood-tracking smartphone applications as a proxy to depressive symptoms  \cite{Canzian2015, nahum2017moodscaler} for various demographic groups, including adolescents \cite{kenny2015copesmart}, university students \cite{lee2018destressify, Wang2014}, clinically diagnosed or high-risk populations \cite{wang2016crosscheck, faherty2017pregnancy}. 
In all these studies, moods are tracked via self-reports that involve user engagement in mobile systems. Regarding the efficacy of self-reported mood tracking, Bakker et al. \cite{bakker2018moodprism} and Birney et al. \cite{birney2016moodhacker} showed through randomized controlled trials that engaging in mobile mood self-monitoring increases emotional self-awareness that results in reducing depressive and anxious symptoms for the clinically depressed population. Besides clinical research, moods in mobile technologies are often associated with more than one behavioral mediator. In one study, Glasgow et al. \cite{glasgow2019travel} showed how travel choices, social ambiance, and destinations are related to moods using mobile experience sampling methodology and GPS location tracking. In another study, Carroll et al. \cite{carroll2013justintime} captured how individuals' moods are associated with emotional eating behaviors. In these studies, moods and mediator behaviors are captured via self-reported, ecological momentary assessment (EMA) responses. 
Recent works in mood and smartphone technology leverage EMA responses with unobtrusive smartphone sensing. One approach is to associate mobile sensor data with mediators. Here, mobile data serves as a descriptive proxy to concurrent activities or environments. Some of them includes physical movements \cite{lathia2017physical, amarasinghe2023multimodal} smartphone usage \cite{cao2017deepmood, zulueta2018keystroke}, social contexts \cite{zhang2018moodbook, chow2017socialAnxiety, mader2024learning}, and sound \cite{rachuri2010emotionsense}. Another approach is to leverage passive sensor data and machine learning to recognize negative moods \cite{faherty2017pregnancy, servia2017mobile} or provide real-time intervention for depressive states \cite{burns2011harnessing}. Such advances in mobile technologies in mood monitoring provide more integrated information, opening the possibility of understanding various human behaviors as holistic events.

\subsection{Eating as a Holistic Event}\label{subsec:eating_as_a_holistic}
Bisogni et al. \cite{bisogni2007dimensions} proposed a contextualized framework for eating and drinking episodes with eight interconnected dimensions after studying why and how people consume food and drinks. Prior studies in mobile sensing have used this framework to model eating and drinking behaviors using sensor data \cite{Santani2018, Biel2018, meegahapola2021one}. \lakmal{This framework is backed by the idea that behavioral and situational factors guide eating.} The eight dimensions are: (1) food or drink: type, source, amount, and how it is consumed; (2) time: chronological, relative experienced; (3) location: general/specific, food access, weather/temperature; (4) activities: nature, salience, active or sedentary; (5) social setting: people present, social processes; (6) mental processes: goals, emotions, and moods; (7) physical condition: nourishment, other status; and (8) recurrence: commonness, frequency, what recurs. \textcolor{new}{Many studies have proposed similar ideas, stating that psychological aspects \cite{Minati2014, Heatherton1991, Herman2005}, activity levels and types \cite{Westerterp2005, Minati2014}, social context \cite{Patel2001, kabir2018factors}, food availability \cite{kabir2018factors, Keenan2018}, and location \cite{kabir2018factors} could affect eating behavior.} 
 As highlighted by these studies, mood is a component that could affect eating behavior to a larger extent. 

Borrowing from such studies, by considering eating as a holistic event with interconnected dimensions, mobile sensing studies have used behaviors and contexts sensed passively to infer various eating behavior-related attributes. Compared to detecting eating episodes (that a user is eating) \cite{Bangamuarachchi2022Sensing}, additional characterizations help understand eating better, hence allowing more personalized and context-aware interventions and feedback \cite{morshed2020real, Meegahapola2021Survey}. Biel et al. \cite{Biel2018} used mobile and wearable sensing modalities to capture everyday eating episodes of a group of 122 Swiss university students and showed that it is possible to infer meal vs. snack eating episodes with an accuracy of around 84\% with mobile diary features. Meegahapola et al. \cite{meegahapola2021one} showed that the self-perceived food consumption level could be inferred with accuracies over 83\% using passive smartphone sensing features. Another study \cite{Meegahapola2020Alone} emphasized the importance of identifying the social context of eating and showed that passive smartphone sensing could be used to infer lonely eating episodes for two student populations in Mexico and Switzerland with accuracies of the range 77\% to 81\%. All these studies leverage the idea that eating is a holistic event, not limited to the food type and amount, as captured in typical mobile food diaries. Moreover, these studies suggest that such sensor-based inferences are useful to reduce user burden in capturing mobile food diaries (because there is a reduced need to capture them from users if they can be inferred) \cite{xu2015more}, providing interventions (because these inferences recognize moments that can be used to provide interventions) \cite{morshed2020real, Meegahapola2021Survey}, and on a larger scale, to understand population-level eating behaviors with passive mobile sensing (which could be useful for behavioral scientists to conduct large-scale studies with reduced self-reporting 
burden on participants) \cite{Bangamuarachchi2022Sensing}. However, even though mood is a key component in better characterizing eating episodes, to the best of our knowledge, mood-while-eating has not been studied in depth in mobile sensing.

\subsection{Smartphone Sensing Inference Personalization}

Prior studies in smartphone sensing have used population-level models for inferences regarding health and well-being-related aspects \cite{Meegahapola2021Survey}. Even though working with sparse and heterogeneous sensing modalities is a challenge, prior work has shown that inferences can be made with reasonable accuracies with such population-level models \cite{servia2017mobile}. However, people are diverse in nature, and so are their behavioral patterns \cite{ferrari2020personalization}. While capturing such diverse behaviors using population-level models has shown promise, prior work has highlighted that more complex tasks need personalization to achieve greater accuracies \cite{taylor2017personalized, bouton2022your}. In addition, Kao et al. \cite{kao2020user} and Jiang et al. \cite{jiang2018towards} have shown that even though population-level models work well in experimental settings, they struggle to generalize well for larger populations due to the heterogeneous nature of individuals. Hence, many commercial mobile health applications have looked into personalization by using user-level models \cite{Meegahapola2020Protecting, likamwa2013moodscope, Bangamuarachchi2022Sensing}. Such models capture the behavioral diversity of people to a greater extent. However, since data from a single user alone is not enough to train a model for that user, hybrid models are also employed (refer to Table \ref{table:keywords_explaination} for the explanation of model types). Both user-level and hybrid models offer personalization to varying extents and have inherent advantages and drawbacks. In this context, there are two main ways of personalizing models \cite{visweswaran2010learning, xu2021leveraging}:

\paragraph{1. Instance-Specific Models} These models would use data and features from the specific instance (a user) in building a model. Re-training or fine-tuning a population-level model by considering individual data (hybrid model) and building user-level models by training models on previously collected data of a user (user-level model) fall under this category. LiKamWa et al. \cite{likamwa2013moodscope} showed that building fully personalized models could increase binary mood inference accuracies compared to population-level models. To personalize, they trained user-level models for participants by training a model using participants' own data and testing on the rest of the data. Similar techniques have been used in ubicomp research in the past \textcolor{new}{\cite{meegahapola2021one, mairittha2020improving, constantinides2018personalized}}. However, such fully personalized models require a single user to provide large amounts of data for model training, and prior work has shown that it is difficult to capture high volumes of data from participants during real-life deployments \cite{kao2020user, jiang2018towards}. LiKamwa et al. showed that for binary mood inference, several days of data were required for user-level models to outperform population-level models. In addition to building user-level models,  our previous work \cite{meegahapola2021one} used neural network model re-training for personalization. The results showed that food consumption level inferences could be made with reasonable accuracies of the range 70\%-80\% using population-level models. However, the results also showed that personalizing by re-training population-level models using a part of the user’s data would increase accuracies by around 1\%-4\%. In this case, since the population-level models already performed well, there is little value in increasing accuracies by small margins with hybrid modeling. It is unknown whether the same technique would work for a different or more complex task where population-level models do not provide reasonable accuracies.

\paragraph{2. Similarity-Based Models.} These models employ similarity distances between instances to form a community of instances for model training, resulting in a hybrid model. Kao et al. \cite{kao2020user} proposed a methodology that utilized collaborative filtering for imputation and clustering techniques to establish fixed user cohorts based on their similarity in sensor data and self-reported health information. For each new user, assignment to a pre-defined cluster is made based on similarity scores, and model training is performed using data from users in the cluster, resulting in improved accuracy for various health-related inferences. A similar technique was used to identify fixed groups of users for mental health severity inferences \cite{palmius2018group}. Additionally, Abdullah et al. \cite{abdullah2012towards} demonstrated that user-clusters can be identified, and models can be trained for clusters instead of individuals for superior performance. Suo et al. \cite{suo2018deep} employed latent distributions obtained from convolutional neural networks (CNNs) to find similarities between users, which were then used to cluster individuals into groups. They showed that their CNN-based approach outperformed other similarity metrics in clustering similar users for disease prediction. In most of the aforementioned studies, a single model is utilized for multiple users if they are assigned to the same cluster. In contrast, our approach does not utilize a set of pre-defined clusters, instead, it identifies a unique community of users for the target user in real-time using cosine similarity of dataset features, resulting in a more personalized outcome for each user. In addition, previous techniques, such as those discussed by Abdullah et al. \cite{abdullah2012towards} assume that a large amount of data is available for each user for personalization, which may not be feasible in tasks such as mood inference. To address this issue, Xu et al. \cite{xu2021leveraging} proposed a collaborative filtering-based approach to detect depression in college students, using majority voting to classify by making a prediction for each dataset feature based on its similarity with a target label. However, it is not clear whether this technique can be used to train more complex models or combined with other machine learning techniques, as the technique is not model-agnostic. Additionally, their technique requires longitudinal data from users and aims to infer a person-level attribute, whereas our approach aims to capture an event-level aspect that may change over time for the same user.

In summary, our technique is model-agnostic and does not assign users to predefined communities. Instead, it identifies a unique community for each target user, with its size adjustable via a tunable threshold parameter. This approach accounts for both interpersonal and intrapersonal variability that may impact model performance. Moreover, our technique effectively tackles the challenge of limited data in the training set by identifying similar users and constructing a community around the selected user.

\section{Dataset Description}\label{sec:mobile_app}

\wageesha{Not many datasets exist to study the intersection of mood and eating behavior with mobile sensing data. Therefore, we utilized two novel datasets collected from two separate studies}. Further details about these datasets are available in our previous works \cite{meegahapola2021one} and \cite{meegahapola2023generalization, busso2025diversityone}, respectively.

\begin{figure*}[t]
\begin{center}

    \begin{subfigure}[t]{0.26\textwidth}
        \centering
        \includegraphics[width=\textwidth]{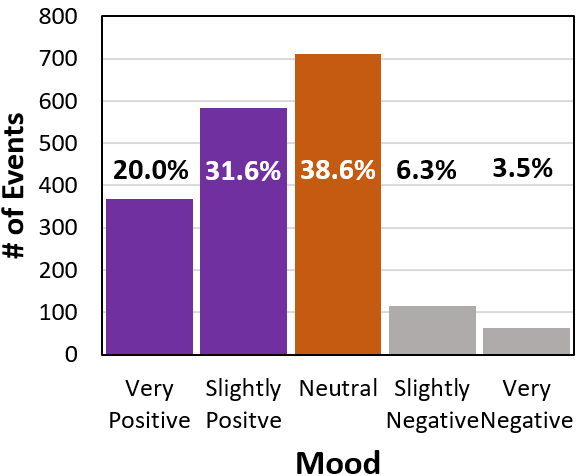}
        \caption{MEX: Original Mood Distribution}
        \label{fig:fi_mood_dis_5_class}
    \end{subfigure}
    \hfill 
    \begin{subfigure}[t]{0.23\textwidth}
        \centering
        \includegraphics[width=\textwidth]{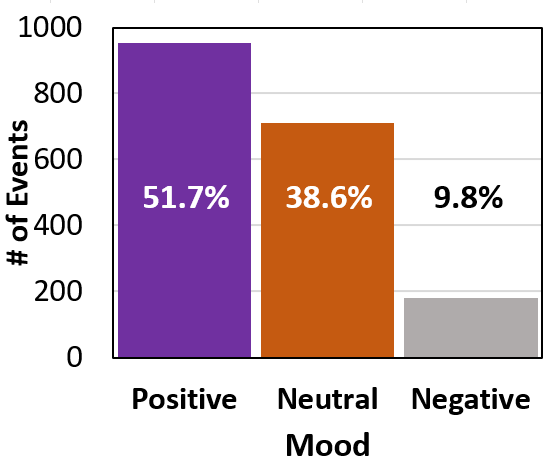}
        \caption{MEX: Three-Class Mood Distribution}
        \label{fig:fi_mood_dis_3_class}
    \end{subfigure}
    \hfill 
    \begin{subfigure}[t]{0.26\textwidth}
        \centering
        \includegraphics[width=\textwidth]{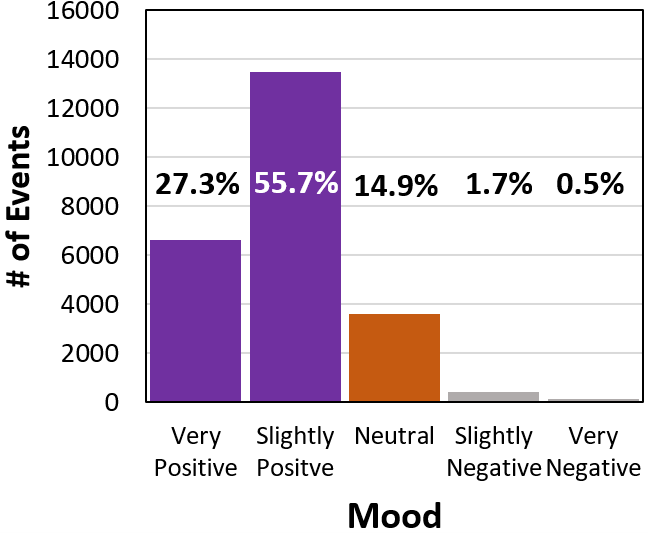}
        \caption{MUL: Original Mood Distribution}
        \label{fig:fi_mood_dis_5_class_new}
    \end{subfigure}
    \hfill 
    \begin{subfigure}[t]{0.23\textwidth}
        \centering
        \includegraphics[width=\textwidth]{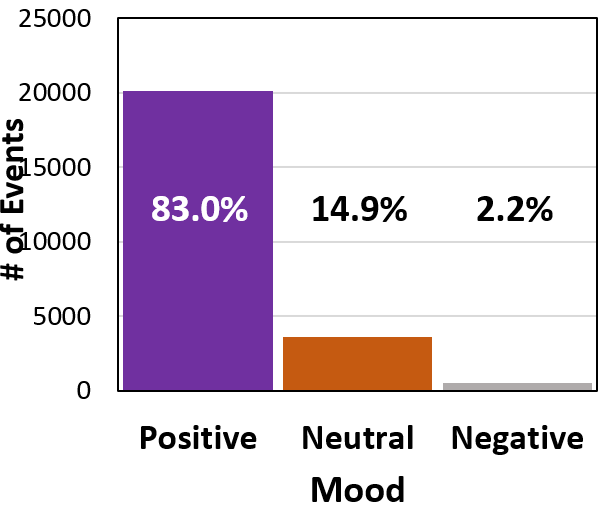}
        \caption{MUL: Three-Class Mood Distribution}
        \label{fig:fi_mood_dis_3_class_new}
    \end{subfigure}

    \caption{Original and Three-Class Mood Distributions of datasets }
    \label{fig:fi_discriptive_analysis_1}
\end{center}
\vspace{-0.2 in}
\end{figure*}

\subsection{{Mexico Dataset (MEX)}}\label{subsec:dataset_mex}

\paragraph{Data collection}\label{subsec:dataset1_collection}

First, we used a dataset about the eating behavior of college students in Mexico \cite{meegahapola2021one}. This dataset has passive smartphone sensing and self-report data regarding the holistic context of eating episodes from 84 college students (56\% women, mean age of 23.4) in Mexico. This dataset is similar to a mobile food diary, but with additional passive sensing features. The objective of collecting the dataset was to show that passive sensing features could be used to infer aspects that could otherwise need self-reports, hence reducing user burden and also enabling context-aware feedback and interventions in mobile food diaries. Therefore, a mobile app called iLog \cite{Zeni2014} was used to collect sensor data and self-reports from participants. The study was carried out between September and December, 2019 among 84 participants. Participants did not suffer from eating disorders and owned an android smartphone. After removing self-reports with missing information and participants with poor sensor data coverage, the final dataset contains a total of 1843 eating episode reports that have matching and sensor data from 61 participants, with an average of around 30 eating reports per participant \wageesha{(standard deviation = 29.39)}. It is also worth noting that sensor data and eating episodes in this study are non-overlapping because reported eating episodes are at least four hours apart. For more information regarding the dataset, please refer to \cite{meegahapola2021one}.

\paragraph{Passive Smartphone Sensing and Self-Reports}\label{subsec:mobileapp_1}

During the deployment phase, participants received three notifications per day and they retrospectively reported the last food intake they had with time of eating to a granularity of half an hour. For each self-report, participants reported the types of food, and more contextual attributes such as semantic location, social context of eating, concurrent activities and mood. All this information provides a more holistic understanding regarding eating episodes from different dimensions in line with the framework by Bisogni et al. \cite{bisogni2007dimensions}. In addition to self-reports, smartphone sensing data were collected from accelerometer, location, battery, screen, and application usage. Further, similar to prior studies that considered a time window around an event to aggregate sensor data to match a self-report \cite{servia2017mobile, Biel2018, Phan2020}, the dataset has been created using a one hour window around the approximate eating time to aggregate sensor data, to match a self-report. Similar to prior work (e.g. 30-120 minutes in \cite{Bae2017}, 120 minutes in \cite{Biel2018}, 60 minutes in \cite{meegahapola2021one}, 60 minutes in \cite{meegahapola2021examining}), a large time window around eating episodes help to capture the behavior and context of participants on and around eating events, hence enabling to understand the overall behavior. For example, this approach would not only capture details about the eating moments, but also, whether participants traveled a long distance before eating (this could be informative of certain patterns, e.g. whether they ate food while being at home or whether they took a bus to and went to some restaurant), whether there was significant physical activity before and after eating (this could suggest that a user walked to a eating location and came back), and whether they used the phone after eating (using youtube to watch videos while eating might induce a different mood as opposed to having to reply to whatsapp and facebook message). \textcolor{new}{Meegahapola et al. \cite{meegahapola2021one} have used a time window of one hour in their study, which is the paper of the original dataset we used and have achieved good results for inferring food consumption level.}
A summary of sensor features is provided in the Appendix, Table~\ref{table:allfeatures}.

\paragraph{Mood-While-Eating Distribution}\label{subsec:mood_distribution_mex}

Figure~\ref{fig:fi_mood_dis_5_class} shows a distribution of 1843 self-reports. The responses were distributed as, very positive (N=369), slightly positive (N=583), neutral (N=711), slightly negative (N=116) and very negative (N=64). After regrouping the responses into three classes, we got the distribution as depicted in Figure~\ref{fig:fi_mood_dis_3_class}. The number of responses considered as negative (N=180) is considerably fewer than the other two moods. As past studies regarding mood have shown \cite{li2020extraction, dubad2018systematic}, these self-reports can be skewed towards positive responses. Hence, this skew was expected, specially for a group of young adults.

\subsection{{Multi-Country Dataset (MUL)}}\label{subsec:dataset_mul}

\paragraph{Data collection}\label{subsec:dataset2_collection} 

The second dataset was initially described in a previous study \cite{meegahapola2023generalization}, and made publicly available in \cite{busso2025diversityone}. The cohort used here, from this larger dataset contains the same type of passive sensing data and self-reported data as the MEX dataset but with more sensing modalities. This dataset was collected from 678 participants (58\% women, mean age of 24.2) in eight countries (China, Denmark, India, Italy, Mexico, Mongolia, Paraguay, UK). \wageesha{Additional demographic information is given in the Appendix, Table \ref{table:MUL_demographocs}}. The dataset was collected between September and November 2020 using the iLog Android mobile application. The objective of this dataset was to collect data from a diverse population of students, which helps to analyze human behavior and context connected with smartphone passive sensing features and self-reports. A total of 329K self-reports were collected from participants, where 24K self-reports were associated with eating episodes\wageesha{, with an average of 92 per user (standard deviation = 58.88)}. In order to get additional information regarding the dataset, please refer to \cite{meegahapola2023generalization} and \cite{busso2025diversityone}.

\paragraph{Passive Smartphone Sensing and Self-Reports}\label{subsec:mobileapp_2}

The passive sensing data were collected throughout the day, and the participants' self-reports were collected many times a day by sending reminder notifications. As self-reported data, users had to provide their current activity, semantic location, social context, and mood. The mobile application collected sensor data such as location, Bluetooth usage, WiFi usage, screen interactions, daily activity level, application usage, and touch events using a multitude of modalities such as accelerometer, gyroscope, location, and magnetic field step counter. The sensor data was aggregated using a 10-minute time window \cite{meegahapola2023generalization}. A shorter time window has been selected in this case in order to prevent self-report/sensor data from getting overlapped. The users were notified many times during the day, using the app's push notifications to complete the survey. Hence, this survey captured the user behavior and context with multimodal data. We use data to identify eating occasions (using the current activity), the mood of the user at the moment, and the aggregated passive sensing data around the eating occasion.

\paragraph{Mood-While-Eating Distribution}\label{subsec:mood_distribution_mul}

Figure~\ref{fig:fi_mood_dis_5_class_new} shows the distribution of 24207 eating self-reports. The distribution consists of very positive (N=6610), slightly positive (N=13475), neutral (N=3597), slightly negative (N=412), and very negative (N=113) mood responses. Figure~\ref{fig:fi_mood_dis_3_class_new} shows the three-class mood distribution after regrouping the five-class mood responses. The data distribution of the MUL dataset is highly skewed towards positive responses compared to the MEX dataset distribution.

\section{Experimental Setup}\label{sec:experiments}

In this section, we discuss the broader experimental setup for research questions. The mood-while-eating inference was performed using several model types, including Random Forest (RF), Naive Bayes (NB), Support Vector Classification (SVC), XGBoost (XGB), and AdaBoost (AB) \cite{rf_inbook, rish2001empirical, natekin2013gradient, chen2016xgboost, schapire2013explaining, noble2006support}. These models were selected based on the dataset’s characteristics, such as its size, tabular nature, and interpretability. To address data imbalance, we applied synthetic minority oversampling (SMOTE) to the training sets to obtain an upsampled dataset. The testing set remained unaltered to reflect real-world scenarios, where oversampling and undersampling are typically not performed. Hyperparameter tuning was conducted using GridSearch. We used scikit-learn \cite{pedregosa2011scikit} along with Python to carry out the experiments. In line with prior work \cite{meegahapola2023generalization}, the three target classes in the experiment were: negative, neutral, and positive. We evaluated multiple models using different techniques with varying degrees of personalization to gain an understanding of how to improve inference accuracy for mood detection during eating episodes. The study was conducted using the following three approaches:

\begin{itemize}[wide, labelwidth=!, labelindent=0 pt]
    \item \textbf{Population-Level Model (PLM):} We employed a population-level, leave-one-user-out strategy \cite{servia2017mobile, meegahapola2023generalization}. In this approach, each target user's data served as the testing set, while the remaining dataset was used for training. The goal was to simulate a scenario in which models were developed using data collected from a population utilizing a mobile health application with mobile sensing and then evaluated on a new target user. Thus, for each selected user, the model was trained on data from the remaining participants and tested on the target user's data. To enhance the robustness of training, for each target user, a randomly sampled subset of 90\% of the population data was used for training, while 30\% of the target user's data was used for testing. This process was repeated five times for each user, and the results were averaged across all users.
    
    \item \textbf{Hybrid Model (HM):} This approach is similar to PLM but includes a portion of the target user's data in the training set, leading to a partially personalized model for that user \cite{likamwa2013moodscope, meegahapola2023generalization}. This method simulates a scenario where the user has already used the sensing app for some time, allowing the model to incorporate personalized data obtained via self-reports from the target user. During testing, newly collected data from the same user was used as the testing set. For each target user, the model was trained using 70\% of that user's randomly sampled data combined with a randomly sampled portion of data from other users. The remaining 30\% of the target user's data was used for testing. The training set did not include the testing data of the selected user. This process was repeated five times for each user, and the results were averaged across all users.
    
    \item \textbf{Community-Based Model (CBM):} The community-based model is based on the similarity-based technique described in Section~\ref{subsec:method_rq3}, where each target user’s community was extracted empirically. To detect the community, different $th$ values were used, and various communities were tested for each user to determine the optimal one. The training set was constructed using a randomly sampled 90\% of the community data for each user, along with 70\% of the target user’s data. The model was then tested on the remaining 30\% of the target user's data. This process was repeated five times for each user, and the results were averaged across all users.  
    
\end{itemize}{}

In the above approaches, we employed random sampling of 90\% of the data to capture the diverse behaviors of users, ensuring robust models across different iterations. Additionally, the 70\%-30\% split of user data is a common practice \cite{joseph2022optimal, nguyen2021influence}, providing a good balance between learning personalized patterns and minimizing overfitting. Furthermore, we conducted experiments for all three approaches using the passive sensing features listed in the Appendix, Table~\ref{table:allfeatures} for MEX and Table~\ref{table:allfeatures_dataset2} for MUL, as obtained from the original datasets. These features were inspired by prior studies \cite{meegahapola2021one, Biel2018, Meegahapola2020Alone, Meegahapola2021Survey}. More details on feature extraction are provided in \cite{meegahapola2021one} and \cite{meegahapola2023generalization}, as well as in the Appendix.

\section{RQ1: Analysis of Generic and Context-specific Models}\label{sec:rq1_methodd_results}

\subsection{{Methods}}\label{subsec:method_rq1}

In order to assess the performance of a generic mood inference model within specific contexts, we utilized the Multi-Country dataset (MUL), which provided information on the concurrent activities being performed by the participants at the time of self-reporting their mood. This dataset included a list of 12 broad activities, such as resting, walking, sports, eating, drinking, studying, and working (see Figure~\ref{fig:fi_distribution_daily_activity_MUL} in the Appendix). By utilizing this dataset, we were able to examine the performance of the mood inference model in different situated contexts as the participants were engaged in various activities. This allows for an examination of the model's performance across a diverse range of settings.

First, we trained a mood inference model using the population-level approach on the MUL dataset. Then, we obtained a breakdown of the mood inference performance for the testing test, for each activity performed by users during the mood report. By examining the mood inference performance across different activities, we aimed to determine whether generic, one-size-fits-all models for mood inference are similarly effective across all situated contexts and whether context-specific models are necessary for eating behavior, as previously studied in the workplace context. Experiments were carried out in both population-level and hybrid modeling approaches, hence examining the partial personalization effect. Experiments were done five times with random sampling, and the results were averaged. Second, considering the total dataset of the size 329K mood reports including the 24K mood-while-eating reports, we conducted an analysis in which we kept the number of training data points from all the activities constant (280K mood reports) and increased the number of mood-while-eating reports in training set from 0 to 15000 (in steps 0, 2500, 5000, 7500, 10000, 12500, 15000). The testing set contained around 20K mood reports captured during other contexts and 2000 mood-while-eating reports. For each step, experiments were done five times with random sampling for training and testing sets, and the results were averaged. Hence, by increasing the number of mood-while-eating reports in the training set and observing the performance of the overall testing set, and in addition, to the 2000 mood-while-eating reports, we examined how the inclusion of data from a specific context affects model performance. Finally, while we reported results for this setup in later sections (considering page limitations and brevity), we also experimented with other different setups, as in, changing the size of the other mood reports in training (e.g., 100K, 150K, 200K, 250K instead of 280K, which is the maximum number of data points we could use for training) and testing sets (e.g., 2.5K, 5K, 10K, 15K instead of 20K, which is closer to the maximum number of data points we could use for testing), which did not yield contrasting conclusions or results.

\subsection{{Results}}\label{subsec:result_rq1}

First, Figure~\ref{fig:context_specific_models} shows that the generic mood inference model (best performing random forest classifier) trained with data from a wide range of everyday life occasions displayed contrasting performance across activities performed while the self-reports were provided. On the one hand, the results indicate that in both population-level and hybrid approaches, activities such as Resting, Walking, and Sports (which correspond to Mood-While-Resting/Walking/Engaging in Sports) showed higher performance than the overall accuracies (PLM: 43.2\%, HM 50.1\%) against a baseline of 33\%. On the other hand, the model performed the worst across Eating, Studying, and Working (which correspond to Mood-While-Eating/Studying/Working). Interestingly, people tend to use phones while resting and walking, while it is not necessarily the case while eating, studying, or working, which are exigent activities. Sport is an exception, where even though it is an exigent activity for which there were not many mood reports, the model still performed decently well. Along this line, prior work has shown that mental well-being during work is an interesting aspect that is worth further investigation \cite{9953824Morshed}. However, as we have discussed previously, such studies do not exist for mood-while-eating. 

Next, Figure~\ref{fig:context_specific_models} shows the effect of having mood reports during eating episodes for the overall performance on the testing set and also the performance of the model only for eating reports in the testing set (mood-while-eating). Results indicate that when no mood-while-eating reports are present in the training set, mood-while-eating performance is 45.6\% and overall performance is 53.3\%. However, when increasing the number of mood-while-eating reports in the training set, performance for eating reports in the testing set increases to 47.8\%. This could be because more representation of mood-while-eating reports in the mood inference model training helped the model generalize better to eating episodes in the testing set. Interestingly, when adding more mood-while-eating reports to the training set, the overall performance declined by around 3\% to 50.1\% before the curve plateaued. This means that even though the number of data points for training the model increased, it did not work well for the overall performance because all the data points came from a similar context (i.e., mood-while-eating). This, in turn, would make the tree-based model learn representations for such contexts better than other contexts, leading to declining performance for other contexts at the expense of increasing performance for mood-while-eating events in the testing set. 

\lakmal{In summary, in response to \textbf{RQ1}, our findings indicate that generic models for mood inference do not exhibit consistent performance across various situated contexts. Similar scenarios where models do not perform for a subset of samples of the distribution have been introduced with terms such as sub-population shift or sub-sample shift \cite{yang2023change, jones2020selective}. Here, what we are observing could be termed as a sub-context shift.} Hence, this result could be attributed to the fact that self-report data may not be equally captured across all daily situations and that the ease of learning mood may vary depending on the context, such as during eating, studying, or working. Furthermore, our analysis revealed that an increase in the representation of mood-while-eating reports in the training set for the mood inference model led to improved performance on mood-while-eating reports in the testing set. \lakmal{However, overall performance declined, highlighting the challenge of generalizing to different situated contexts. These results suggest the need for context-specific models or better domain generalization techniques for mood inference in order to improve the accuracy of continuous mood tracking in mobile food diaries and health applications. Hence, in the next RQs, we explore the first approach of building context-specific mood inference models for eating occasions.}

\begin{figure*}[t]
\begin{center}

    \begin{subfigure}[t]{0.65\textwidth}
        \centering
        \includegraphics[width=\textwidth]{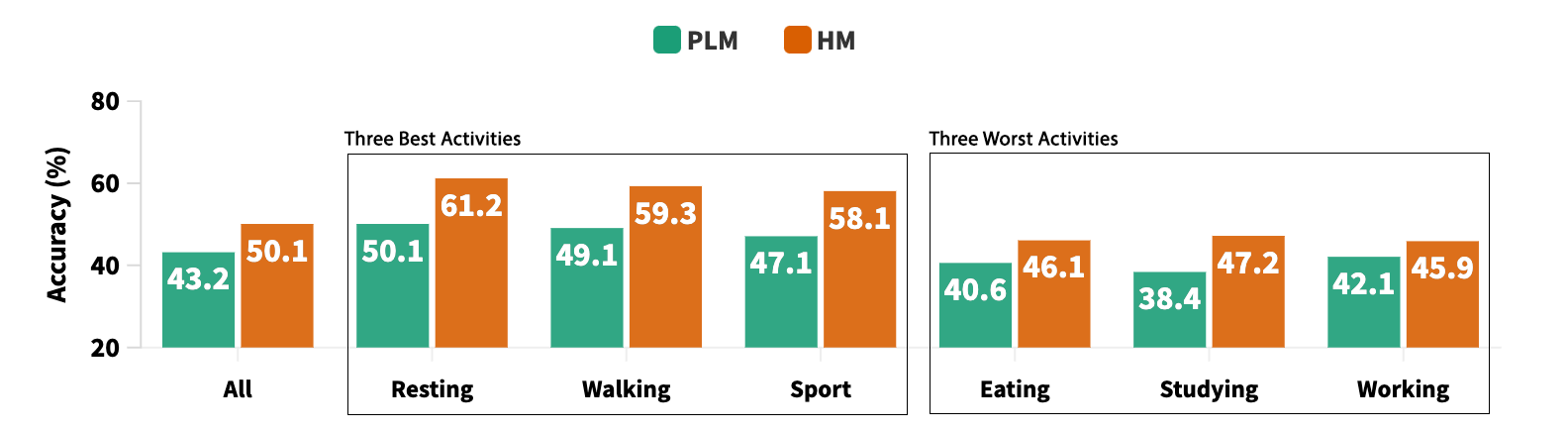}
        \caption{}
        \label{fig:context_specific_models}
    \end{subfigure}
    \hfill 
    \begin{subfigure}[t]{0.34\textwidth}
        \centering
        \includegraphics[width=\textwidth]{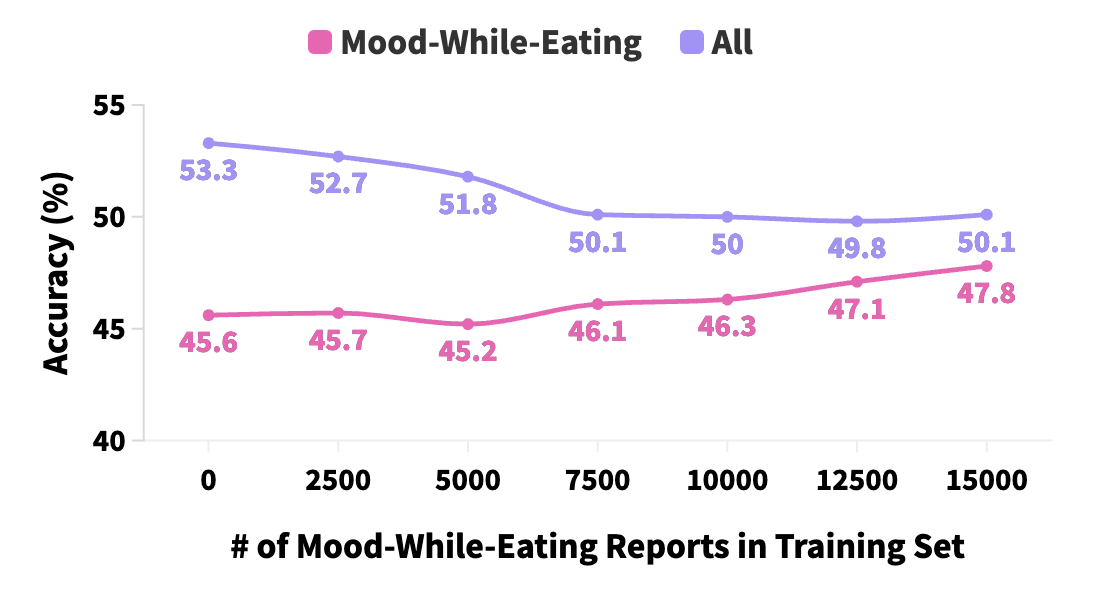}
        \caption{}
        \label{fig:changing_eating_events}
    \end{subfigure}
    
    \caption{(a) This shows the context-specific accuracies of the generic mood inference model trained for MUL for approaches PLM and HM. (b) This shows the overall mood inference accuracy (All) and accuracy for eating episodes (mood-while-eating) when the number of mood-while-eating reports in the training set changes, with HMs for MUL.}
    \label{fig:context_specific_models}

\end{center}
\vspace{-0.2 in}
\end{figure*}

\section{RQ2: Mood-While-Eating Inference}\label{sec:rq2_methodd_results}

\subsection{{Methods}}\label{subsec:method_rq2}

\lakmal{For this section, we utilized both datasets. In the MEX dataset, we used the entire dataset since all mood reports were related to eating episodes, thus providing data for mood-while-eating. However, in the MUL dataset, which contains over 329K mood reports, with the majority corresponding to various activities, we only used mood reports associated with eating episodes to examine this research question. We initially trained population-level and hybrid models to assess the effectiveness of mood inference. Experiments were conducted using both approaches with multiple model types (RF, NB, SVC, XGB, AB) across both datasets. Additionally, we discuss the reasons why we were unable to examine and train user-level models for mood-while-eating inference in settings with limited data for each individual. Therefore, we identify the need for a more effective personalization strategy in limited data scenarios.}

\subsection{{Results}}\label{subsec:result_rq2}

\subsubsection{{Population-Level and Hybrid Model Results}}\label{subsubsec:inference_results_plm}

The results of the PLM and HM techniques are summarized in Table~\ref{tab:inference_results_pop_loo}. The highest accuracy, F1 score, and AUC-ROC value obtained for the MEX dataset were 38.6\%, 38.6\%, and 0.51, respectively, obtained by different model types. For the MUL dataset, the highest accuracy was 83.1\%, and the highest F1 Score was 83.1\% with SVC and XGB models—the highest AUC-ROC score was 0.51 with RF and NB model types. Considering both MEX and MUL performances, NB, SVC, and XGB did not perform well with both datasets. In summary, the PLM approach did not perform well for the three-class inference task regardless of the data or model type. This could be because of the averaging effect in models, as previous work has pointed out \cite{pratap2019accuracy, kao2020user, xu2021leveraging}. In addition, the best accuracy for the MEX dataset from the HM approach is 54.6\%, and for the MUL dataset, the accuracy is 83.7\%.

\begin{table}[t]
        \small
        \centering
        \caption{Accuracy (\={A}) [with standard deviation (A$_{\sigma}$)], F1-Score (F1) [with standard deviation (F1$_{\sigma}$)] and AUC-ROC value (AUC)  [with standard deviation (AUC$_{\sigma}$)], calculated using five models for the PLM and HM of mood inference task: \={A} (A$_{\sigma}$), F1, AUC}
        \resizebox{0.99\textwidth}{!}{%
        \begin{tabular}{l l l l l l l}

        \rowcolor{gray!10}
        \textbf{}& 
        \textbf{} &
        \textbf{} &
        \textbf{PLM}&
        \textbf{}&
        \textbf{}&
        \textbf{HM} 
        \\

        
        \arrayrulecolor{gray!15}
        \cmidrule{2-6}

        \textbf{Dataset (\# of Features)}& 
        \textbf{RF} &
        \textbf{NB} &
        \textbf{SVC}&
        \textbf{XGB}&
        \textbf{AB} &
        \textbf{RF}
        \\
        
        \textbf{}& 
        \textcolor{new}{\={A}, F1, AUC}&
        \textcolor{new}{\={A}, F1, AUC}&
        \textcolor{new}{\={A}, F1, AUC}&
        \textcolor{new}{\={A}, F1, AUC}&
        \textcolor{new}{\={A}, F1, AUC}&
        \textcolor{new}{\={A}, F1, AUC}
        \\

        \textbf{}& 
        \textcolor{new}{(A$_{\sigma}$), (F1$_{\sigma}$), (AUC$_{\sigma}$)}&
        \textcolor{new}{(A$_{\sigma}$), (F1$_{\sigma}$), (AUC$_{\sigma}$)}&
        \textcolor{new}{(A$_{\sigma}$), (F1$_{\sigma}$), (AUC$_{\sigma}$)}&
        \textcolor{new}{(A$_{\sigma}$), (F1$_{\sigma}$), (AUC$_{\sigma}$)}&
        \textcolor{new}{(A$_{\sigma}$), (F1$_{\sigma}$), (AUC$_{\sigma}$)}&
        \textcolor{new}{(A$_{\sigma}$), (F1$_{\sigma}$), (AUC$_{\sigma}$)}
        \\

        \arrayrulecolor{Gray}
\hline 
        
        MEX (40) & 
        24.1, 25.1, 0.47 & 
        38.6, 38.6, 0.48 & 
        17.6, 17.8, 0.51 & 
        18.6, 18.6, 0.48 & 
        26.3, 26.3, 0.48 &
        54.6, 55.1, 0.43  
        \\

        & 
        (8.17), (8.05), (0.02)  & 
        (5.57), (5.56), (0.02) & 
        (6.85), (6.85), (0.02) & 
        (9.2), (9.29), (0.03) & 
        (10.6), (10.5), (0.04) &
        (15.5), (16.5), (0.20)  
        \\
        

        \arrayrulecolor{Gray}
\hline 
        
        MUL (114) &
        82.4, 76.8, 0.51 & 
        6.7, 6.7, 0.51 & 
        83.1, 83.1, 0.50 & 
        83.1, 83.1, 0.50 & 
        82.9, 82.9, 0.49 &
        83.7, 79.4, 0.44 
        \\

        &
        (0.56), (0.40), (0.03)  & 
        (1.67), (1.67), (0.02) & 
        (0.0), (0.0), (0.03) & 
        (0.0), (0.0), (0.04) & 
        (0.39), (0.39), (0.03) &
        (4.87), (6.19), (0.29)  
        \\

        \hline 
        
        \end{tabular}}
        \label{tab:inference_results_pop_loo}

\end{table}

\subsubsection{User-Level Results}\label{subsubsec:ulm}

\begin{wrapfigure}{r}{0.3\textwidth}
  \begin{center}
    \centering
    \includegraphics[width=0.9\linewidth]{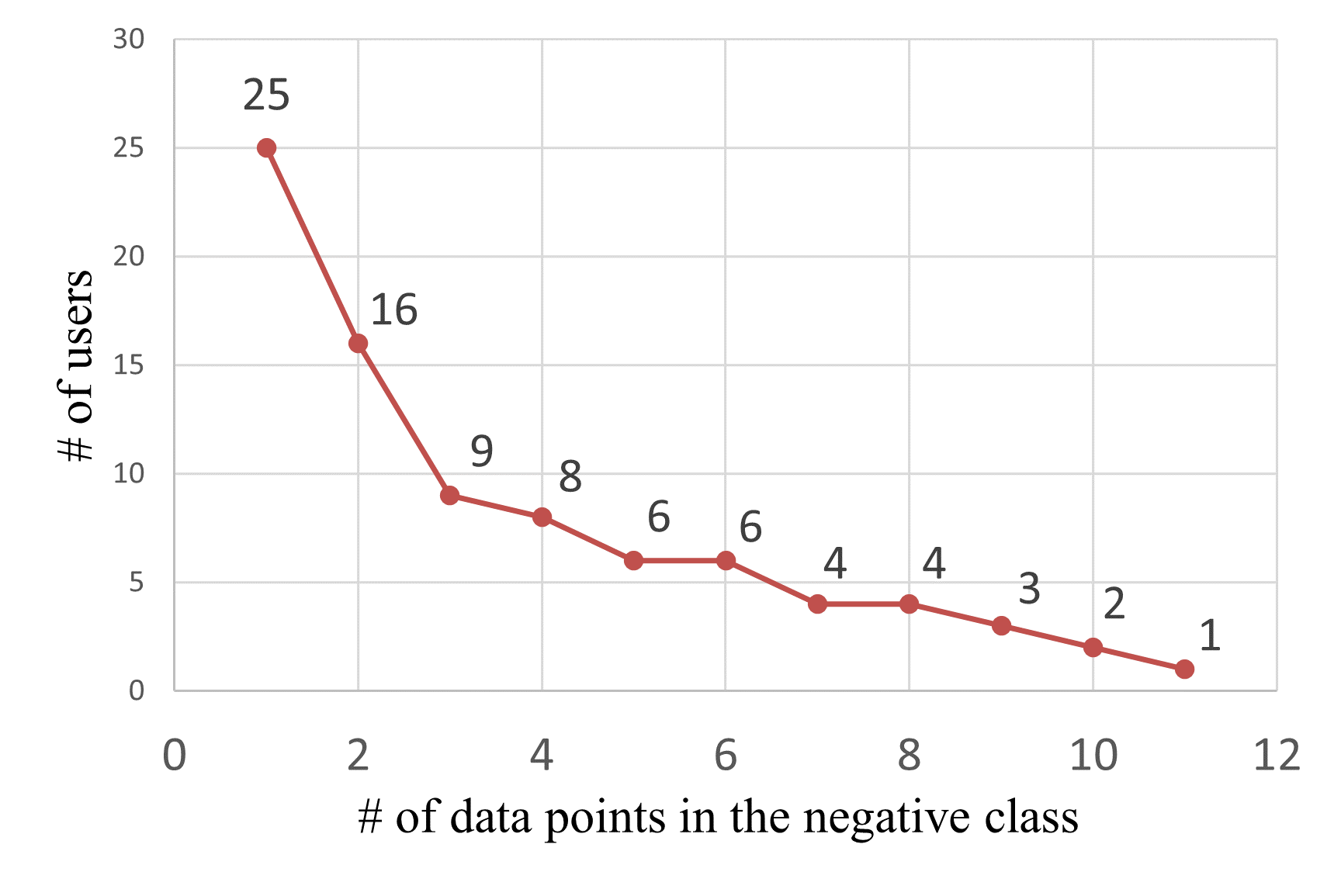}
    \caption{MEX dataset: Distribution change in no. of users with the increase of no. of data points in the negative class.}
    \label{fig:ulm_users}
  \end{center}
\end{wrapfigure}

Even though we did not report results for user-level models (ULM), we conducted experiments with them for both datasets. ULMs are models that only use the target users' data in both training and testing with a 70:30 split (refer Table \ref{table:keywords_explaination} for explanation). However, the number of users for whom experiments could be carried out was fairly low because there was a low number of users with enough data to be split into training and testing sets. For example, in the MEX dataset, if we consider the number of classes represented in the dataset for each user, there are four users with data from only one class (only positive or only neutral), 32 users with data from only two classes (only two classes out of positive, negative, and neutral), and 25 users with data from all three classes (all positive, negative, and neutral). Hence, a generic ULM can only be built for these 25 users. However, as shown in Figure~\ref{fig:ulm_users}, the number of users with a high number of negative class labels decreases significantly when the number of labels is increased. For example, there are 16 users with at least two negative class labels, and it goes down to 4 users with at-least eight negative class labels. However, when data is split into training and testing, there might not be enough labels. Moreover, in the MUL dataset, the same issue can be identified, where there was a lesser percentage of negative class labels compared to the MEX dataset, which makes it more difficult to build user-level models using MUL dataset. Due to these reasons, ULMs can not be trained for a majority of users under reasonable assumptions. For the purpose of understanding, with an arbitrary negative label count of 6, we could train ULMs with the MEX dataset for six users with 70:30 training and testing splits and achieve an accuracy of 73.5\% with passive sensing modalities.

Overall, these results of \textbf{RQ2} suggest that, even though model types such as population-level, hybrid, and user-level can be used to infer mood-while-eating, either they lack performance or the model cannot be used for a large number of users due to lack of data.

\section{RQ3: Community-Based Model Personalization Approach}\label{sec:rq3_methodd_results}

\subsection{{Methods}}\label{subsec:method_rq3}

\begin{figure*}[t]
\begin{center}
    \begin{minipage}[t]{\textwidth}
        \centering
        \includegraphics[width=0.9\textwidth]{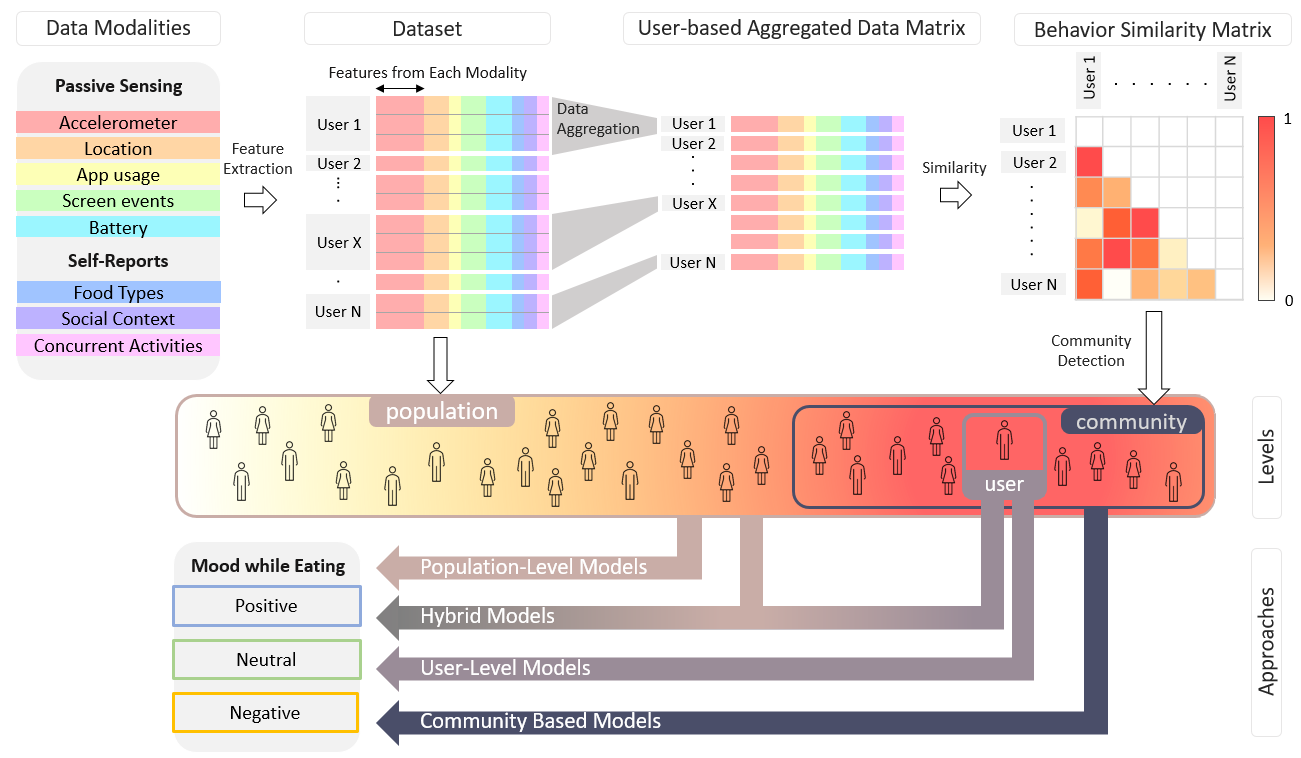}
        \caption{High-Level Architectural View of the Community-based Personalization Approach}
        \label{fig:architecture}
    \end{minipage}
\end{center}
\vspace{0.035 in}
\end{figure*}

When a large population of users is considered, there can be users who are different from each other as well as users with certain similarities. With the proposed technique, we aim to discover similar users for a target user and categorize them into a group (i.e., the community of the target user) in order to increase the number of data points in the training dataset when compared to training a user-level model. This is because gathering a large amount of data from a single user can be time-consuming, and it could allow higher accuracies as compared to population-level models \cite{kao2020user, jiang2018towards}. As mentioned previously, the usage limitations of the user-level models because of the number of data points in the negative class and why this personalization technique is proposed for this specific inference task is discussed in more detail in Section~\ref{subsubsec:ulm} and Section~\ref{subsubsec:inference_results_compare}. 
A high-level architectural view of the approach we used to infer mood-while-eating is shown in Figure~\ref{fig:architecture}. The approach consists of a user-level data aggregation method and user similarity matrix calculation as described in Section~\ref{sub:user_similarity_detection} and a community detection mechanism described as described in Section~\ref{sub:community_detection}. The purpose of the community-based personalization approach is to discover similar users for a given target user based on their behavioral and contextual data.

\subsubsection{Obtaining the User Similarity Matrix}\label{sub:user_similarity_detection}

\paragraph{User-based Data Aggregation}\label{sub:user_based_data_aggregation}

First, to quantify the similarity between users, we use a mean-based data aggregation method similar to \cite{mitani2006local}. As given in Algorithm~\ref{alg:algorithm_user_based_data_aggregation} in the Appendix, for each user, we first filter out the data points of that user from the dataset ($D_{u}$) using the user id. Then we calculate the mean for each feature and generate a feature vector ($u_{aggr}$) for each user, by considering all the available data points. Hence, the feature vector of each user would have one value for all features in the dataset. Then, the feature vectors were included in the user-based aggregation matrix ($U_{aggr}$). This matrix provides an overview of all the users in the dataset and a summary of their features. In addition, this representation allows comparing users based on their feature vector.

\paragraph{User Similarity Matrix}\label{sub:user_similarity_matrix}

By using $U_{aggr}$ matrix, the next step is to calculate similarities between users. When we consider two vectors, the similarity between them can be calculated with the cosine similarity \cite{li2019improved}. Consider two aggregated user vectors $u1_{aggr}$ and $u2_{aggr}$ of randomly selected user1 and user2, respectively. Then we can find the similarity between those two users with Formula ~\ref{eqn:eqn_cosine} \cite{li2008mining}, where $i$ is the $i^{th}$ column index of the $u_{aggr}$ (i.e. $i^{th}$ feature of $F$).

\begin{equation}\label{eqn:eqn_cosine}
Sim_{cosine} = cos(\Theta) = \frac{A\cdot B}{\left \| A \right \|\left \| B \right \|} = \frac{\sum_{i=1}^{n} A_{i}\times B_{i}}{\sqrt{\sum_{i=1}^{n}(A_{i}^{2})}\times\sqrt{\sum_{i=1}^{n}(B_{i}^{2})}} = \frac{\sum_{i=1}^{|F|} u1_{aggr_{(i)}}\times u2_{aggr_{(i)}}}{\sqrt{\sum_{i=1}^{|F|}(u1_{aggr_{(i)}})^{2}}\times\sqrt{\sum_{i=1}^{|F|}(u2_{aggr_{(i)}})^{2}}}
\end{equation}

This similarity value between the two users suggests how close they are based on their smartphone sensing features and self-reports. We calculate this metric for all user pairs and represent it in a matrix of dimensions $|U|^{2}$. For each user pair, this value suggests how similar they are, where similarity increases when going from 0 to 1 (this is after normalizing the original -1 to +1 scale). Similar users to a target user can be selected using this metric.

\subsubsection{Community Detection}\label{sub:community_detection}

In this section, we discuss the approach of threshold selection, where we select different community sizes based on a tunable threshold value and community detection.

\paragraph{Threshold Selection}\label{sub:threshold_selection}

The similarity values calculated in the previous step were considered as threshold (th) values, which can be used to filter users similar to the target user. Depending on the value of the threshold, the size of the community would differ. For example, if $th = 0.85$, we take all the users who have similarity values equal to or greater than 0.85 with the target user as the community. Hence having the threshold at zero keeps all the other users in the community (similar to a population-level model), and we can increase the threshold to reduce the number of users in the community. While this makes the community much smaller, it also makes it much more similar to the target user. On the other hand, increasing the threshold helps to remove some users from the training set, who could have been noisy otherwise. Increasing the $th$ would affect the resulting dataset in two ways. First, it would reduce the number of users in the community, hence leading to comparatively smaller datasets for training models. Second, it is made sure that the selected data points are coming from similar users. Hence, depending on the chosen $th$, it could increase or decrease the size of the resulting dataset, hence affecting the accuracies of any model, trained on the dataset. In our experiment, we used a range of $th$ values to obtain accuracies for all users.

\paragraph{Community Detection}\label{sub:user_community_detection}

Depending on $th$, for each user, the number of similar users ($|U_{u_{t}}|$) might vary. As stated in the Algorithm~\ref{alg:algorithm_user_community_detection} in the Appendix, $Sim_{u_{t}}$ for each user is a vector with the dimensions of $1 \times (|U|-1)$, which contains the similarity matrix calculated for target user $u_{t}$, with other users in the dataset (i.e. $U \textbackslash \{u_{t}\}$). When iterating through similarity values for the target user, $Corr_{u}$ which is a similarity value of the target user with another user, is compared against $th$ to decide whether the other user is included in the community of the target user. After the community is detected, training and testing of personalized models can be done using the community dataset.

\subsection{{Results}}\label{subsec:result_rq3}

\subsubsection{{Community-Based Model Results with Changing Thresholds}}\label{subsubsec:inference_results_ulcm} 

\begin{figure*}[t]
\begin{center}

    \begin{subfigure}[t]{0.49\textwidth}
        \centering
        \includegraphics[width=\textwidth]{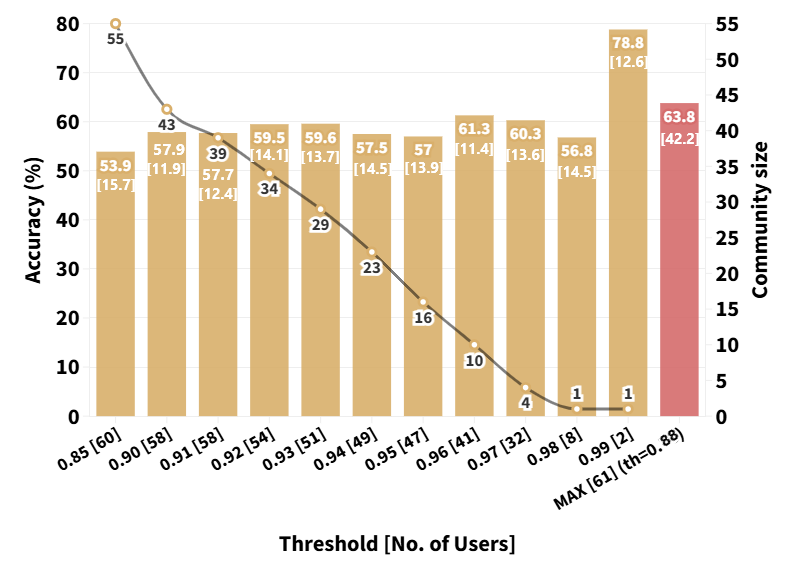}
        \caption{\wageesha{MEX dataset}}
        \label{fig:clm_th_mex}
    \end{subfigure}
    \begin{subfigure}[t]{0.49\textwidth}
        \centering
        \includegraphics[width=\textwidth]{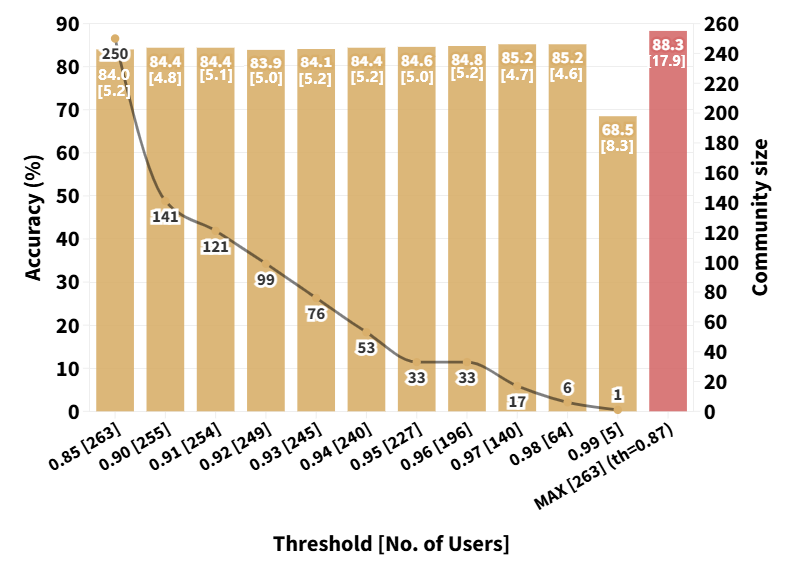}
        \caption{\wageesha{MUL dataset}}
        \label{fig:clm_th_mul}
    \end{subfigure}
  
    \vspace{-0.1 in}
    \caption{Mean accuracy values (column values in the graphs) [with standard deviations] calculated using the random forest for CBM for multiple threshold values [No. of Users] and averaged community sizes (line values in the graphs)}
    \label{fig:clm_threshold_vs_acc}
\end{center}
\vspace{-0.2 in}
\end{figure*}

Figure~\ref{fig:clm_threshold_vs_acc} summarized the results of the CBM approach with MEX and MUL datasets.  Since RF models performed relatively well with the CBM approach and considering space limitations, we only discuss results obtained from RFs from here onwards. Both Figure~\ref{fig:clm_th_mex} and Figure~\ref{fig:clm_th_mul} show results of models when the threshold ($th$) value is increased from 0.85 to 0.99. Even though we obtained results for other threshold values ranging from 0 to 0.99, we only included threshold values that showed high variations in terms of accuracies, average community size, and the number of people to which the inference can be applied [No. of Users]. The point to note is that in these cases, the same threshold has been used on all users, which is not ideal. For example, for the MEX dataset, accuracy increases to a maximum of 78.8\% (at $th = 0.99$).  This is mainly because as $th$ increases, the similarity of users increases, hence leading to more similar data samples for a target user. This could increase model performance. 
Even though there is no noticeable increase in accuracy in the MUL dataset with the increase in threshold values as the MEX dataset, similar behavior can be identified. However, in the MUL dataset, the accuracy decreases after $th = 0.98$. This is because as the threshold increases, the number of people in the community decreases, leading to a low number of samples for training models for a target user.
The comparatively higher accuracies observed in the MUL dataset can potentially be explained by comparison to the MEX dataset (positive 51.7\%, negative 9.8\%), the MUL dataset has a more unbalanced dataset where the percentage of positive labels is 83\%, and the percentage of negative labels is 2.2\%. Hence, when building the models, this leads to more positive labels for the selected users included in the training set, which aligns with the testing set. 
In addition, we mention the number of users for which models can be trained using the approach for different thresholds. While a  $th=0.85$ allows creating models for 60 users in the MEX dataset,  $th=0.99$ only allows models for two users. The reason for this again is that for many users, there is no community at such high thresholds.

In the rightmost column (MAX), we use different thresholds for different users, and the average threshold values for all the users are given within brackets (e.g., th=0.88). We empirically found the ideal threshold for each target user that yields the highest accuracy for the inference. Since the community size decreases when we increase the threshold, in order to build a model for a target user, there should be users with similarity values greater than the given threshold. MAX column summarizes the average results for such different thresholds. These are the best accuracies that we could expect when CBMs are deployed. For example, the accuracy value 63.8\% obtained for the MEX dataset (rightmost column) represents an average of maximum accuracies obtained for all the users. The 0.88 threshold value represents the average of threshold values associated with those accuracy values for each user. For the MUL dataset, the maximum accuracy obtained was 88.3\%, with an average threshold of 0.87. Therefore, with CBMs, sensors performed reasonably well, showing the potential of making mood inferences for eating occasions just by using sensor data. Moreover, while such adaptive thresholds increased the overall accuracies, they also allowed the creation of CBMs for all 61 users in the MEX dataset and 263 users in the MUL dataset.

\subsubsection{{Comparison of Approaches}}\label{subsubsec:inference_results_compare}

\begin{figure*}[t]
\begin{center}

    \begin{subfigure}[t]{0.3\textwidth}
        \centering
        \includegraphics[width=\textwidth]{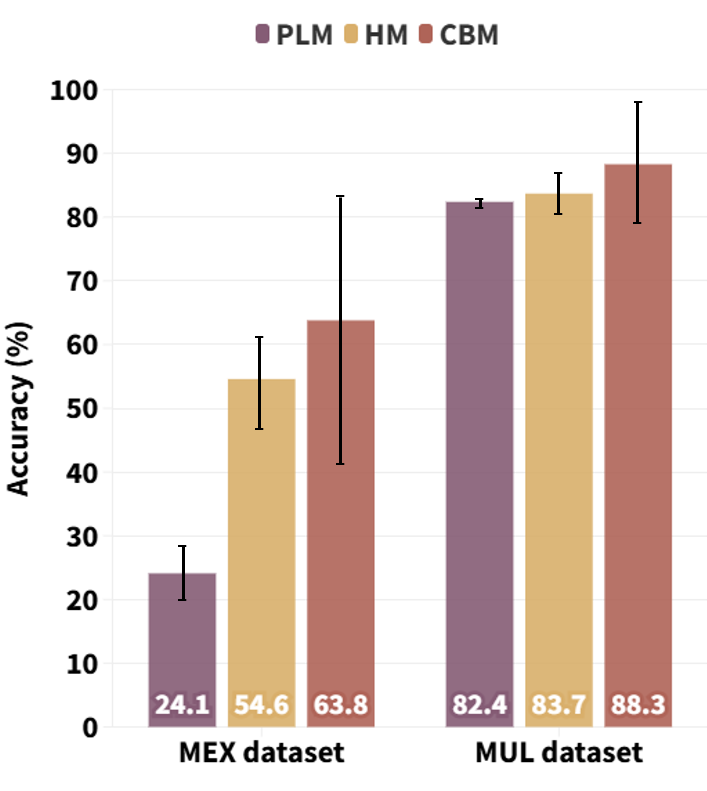}
        \caption{\wageesha{Accuracy}}
        \label{fig:plm_hm_ulcm_comparison_accuracy}
    \end{subfigure}
    \hfill 
    \begin{subfigure}[t]{0.3\textwidth}
        \centering
        \includegraphics[width=\textwidth]{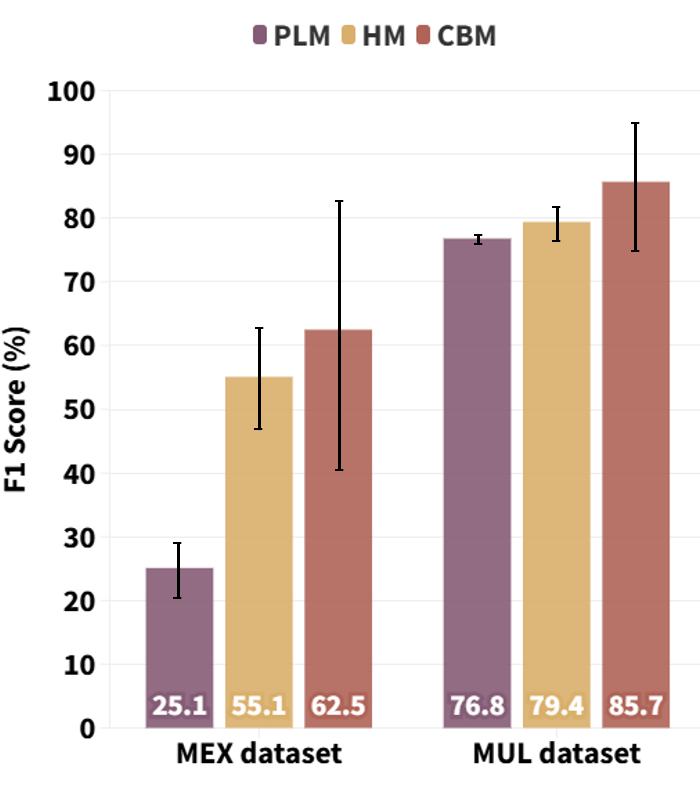}
        \caption{\wageesha{F1 Score}}
        \label{fig:plm_hm_ulcm_comparison_f1}
    \end{subfigure}
    \hfill 
    \begin{subfigure}[t]{0.3\textwidth}
        \centering
        \includegraphics[width=\textwidth]{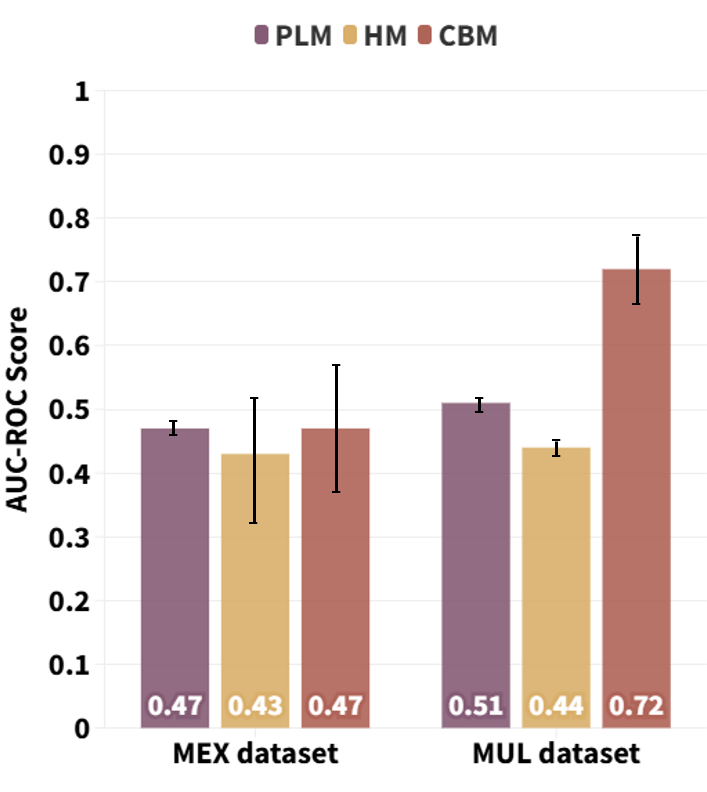}
        \caption{AUC-ROC Score}
        \label{fig:plm_hm_ulcm_comparison_auc}
    \end{subfigure}
  
    \vspace{-0.1 in}
    \caption{Mean accuracies, F1 Scores and AUC-ROC scores [with error bars] calculated using the random forest for the PLM and HM of mood detection task for all the users and Maximum mean values for CBM calculated for all the users.}
    \label{fig:plm_hm_ulcm_comparison}
\end{center}
\vspace{-0.2 in}
\end{figure*}

The results of the population-level, hybrid, and community-based approaches, obtained with Random Forests (RFs) for both the MEX and MUL datasets, are depicted in Figure~\ref{fig:plm_hm_ulcm_comparison_accuracy}. The F1 scores and AUC-ROC scores for the three model types across both datasets are presented in Figure~\ref{fig:plm_hm_ulcm_comparison_f1} and Figure~\ref{fig:plm_hm_ulcm_comparison_auc}, respectively.

In real-world mHealth applications, it is difficult to capture large amounts of data from users \cite{kao2020user, jiang2018towards}. This is the main drawback of user-level models (i.e. the model is trained and tested by splitting the target user's data into two splits), even though they provide reasonably high accuracies. The MEX dataset was collected from the users for a reasonably long time period (close to 30 days). However, the number of collected data points was not sufficient to build a user-level model for many users because of the lack of negative class labels. The same goes with MUL dataset where the number of negative class labels in the dataset is not sufficient to build a 3-class model for most of the users. Additionally, there is a possibility of over-fitting user-level models to the target user when training the model using only the user's data because the model does not even account for intra-user variability, let alone inter-user variability. Hence, if the user's behavior changes with time, there is a chance of not being able to predict the mood correctly with the previously trained user-level models. Furthermore, unlike user-level models, the machine learning model that had been built for each user shows robustness against behavioral fluctuations because the training split contains data from different users, which helps to avoid over-fitting. Hence, the CBM approach has the ability to capture both the inter-personal and intra-personal diversity of a target user. Moreover, with CBM approach the community of the selected user would be adjusted to the user's behavior change with the context around them.

\begin{figure*}[t]
\begin{center}

    \begin{subfigure}[t]{0.38\textwidth}
        \centering
        \includegraphics[width=\textwidth]{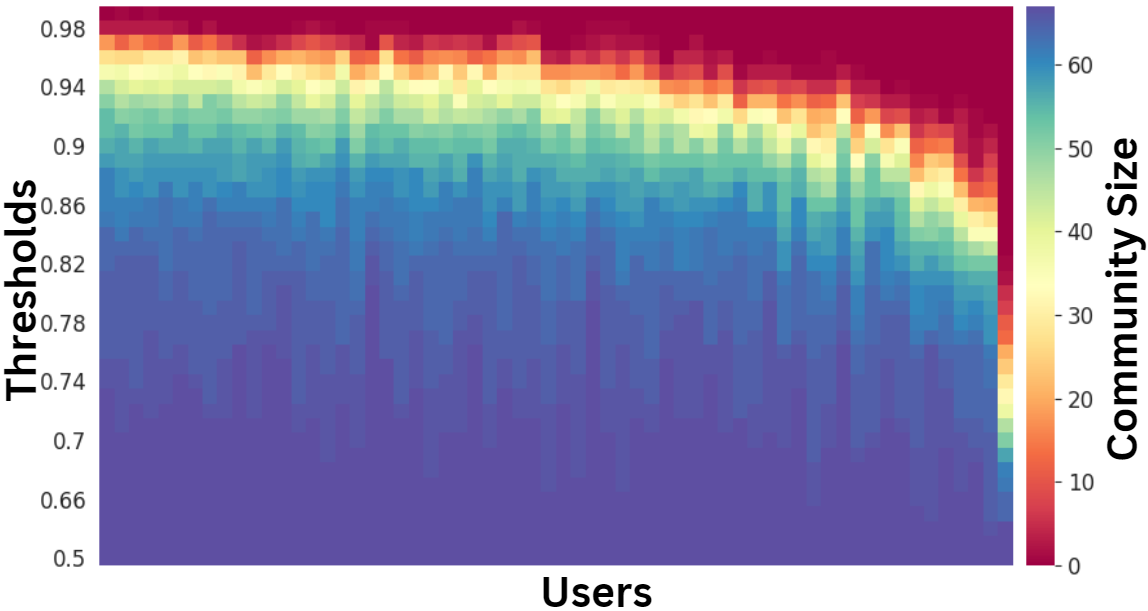}
        \caption{Community size of users with threshold.}
        \label{fig:fi_heat}
    \end{subfigure}
    \hfill 
    \begin{subfigure}[t]{0.30\textwidth}
        \centering
        \includegraphics[width=\textwidth]{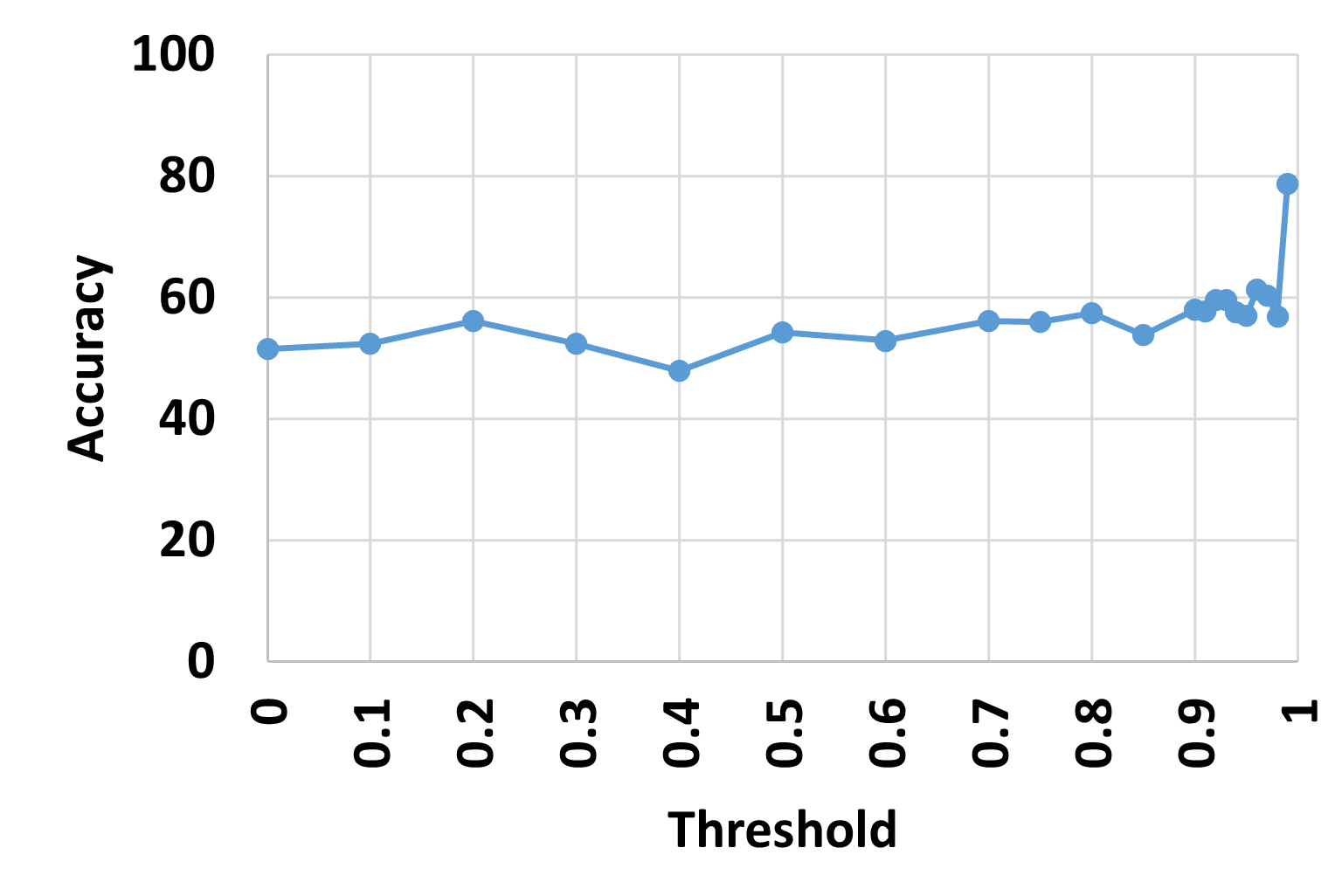}
        \caption{Random Forest Accuracy with threshold.}
        \label{fig:fi_acc}
    \end{subfigure}
    \hfill 
    \begin{subfigure}[t]{0.30\textwidth}
        \centering
        \includegraphics[width=\textwidth]{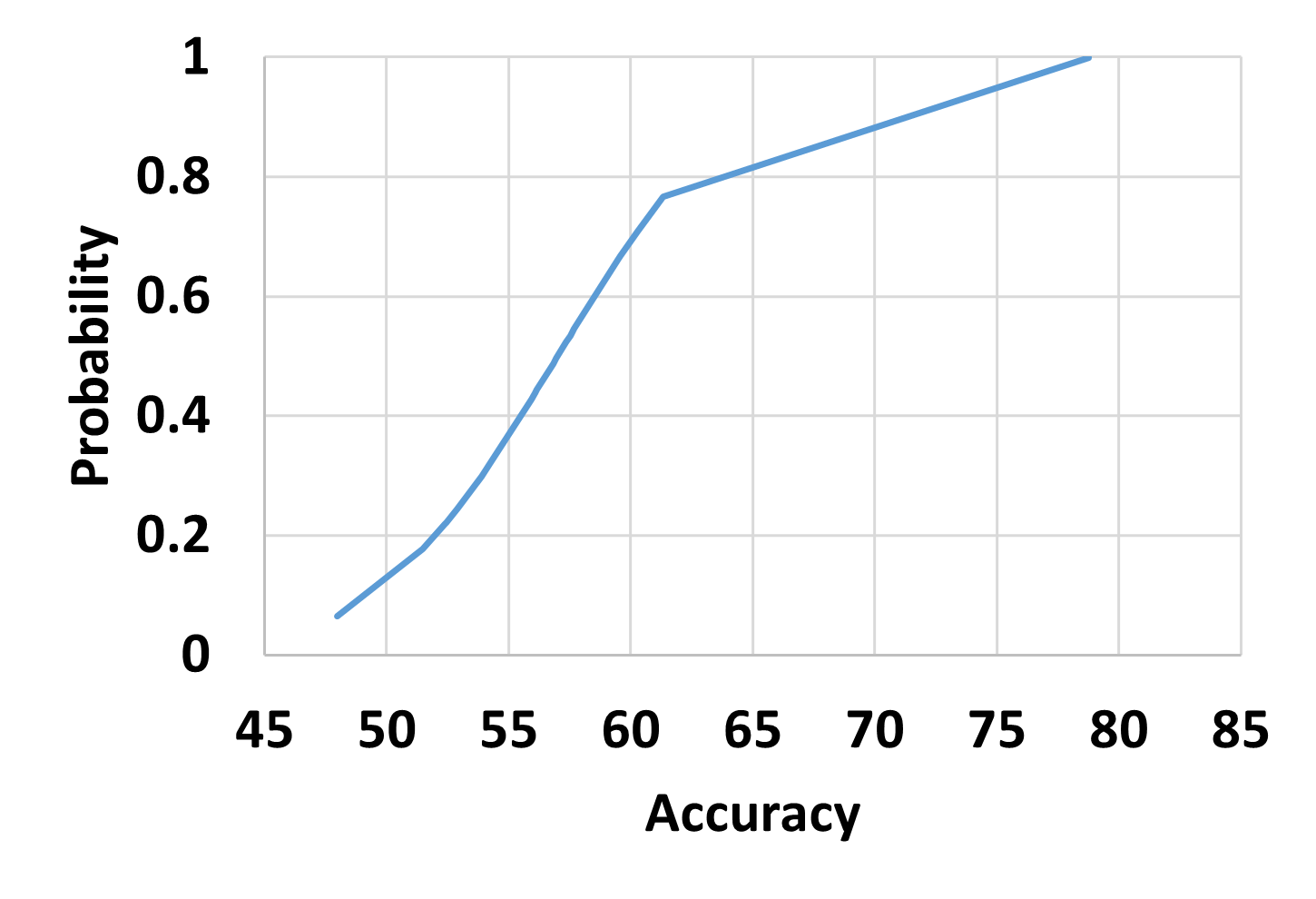}
        \caption{CDF of Accuracy}
        \label{fig:fi_pdf_cdf}
    \end{subfigure}
    \hfill 
    
    \caption{Distributions of CBM using MEX dataset: (a) How the community size of each user changes with threshold $th$; (b) How the mean accuracy of the Random Forest model changes with threshold $th$; (c) Cumulative Distribution Function (CDF) of Accuracy.}
    \label{fig:feature_importance_pers2}
\end{center}
\vspace{-0.2 in}
\end{figure*}

As shown in Figure~\ref{fig:clm_threshold_vs_acc}, we observed that the average community sizes decrease with the increase of threshold value. It can be observed that each user has different threshold values and community sizes. To get an overall idea of the community size variability with multiple threshold values, a heat map generated using the MEX dataset is given in Figure~\ref{fig:fi_heat}. It shows how the community size is reduced when the threshold is increased. Figure~\ref{fig:fi_acc} shows the average of accuracies obtained for all the users. It clearly depicts that the accuracy slightly increases with the threshold. This concludes that for any user, when the threshold is increased, the community size of that user decreases and the accuracy might increase. However, after high thresholds, the accuracy drops off again because of lack of members in user communities. A cumulative density function (CDF) figure for maximum accuracies is included in Figure~\ref{fig:fi_pdf_cdf}. Please note that considering the space limitation we have only included the distributions of MEX dataset.

\subsubsection{{Feature Importance for Mood Detection}}

\begin{figure*}[t]
\begin{center}
    \begin{minipage}[t]{0.5\textwidth}
        \centering
        \includegraphics[width=0.99\textwidth]{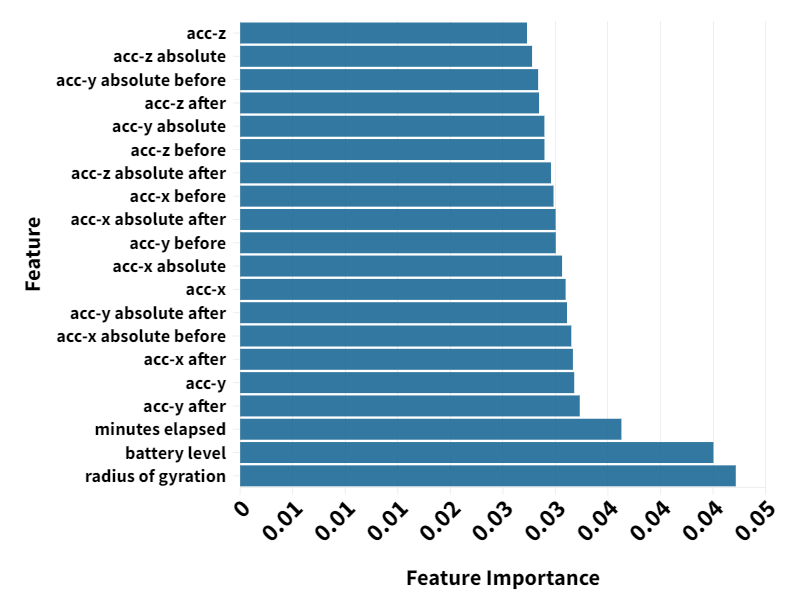}
        \caption{\wageesha{Gini Feature Importance values of top 20 features generated using Random Forest for CBM of MEX dataset}}
        \label{fig:fi_clm_mex}
    \end{minipage}
    \hfill 
    \begin{minipage}[t]{0.49\textwidth}
        \centering
        \includegraphics[width=0.99\linewidth]{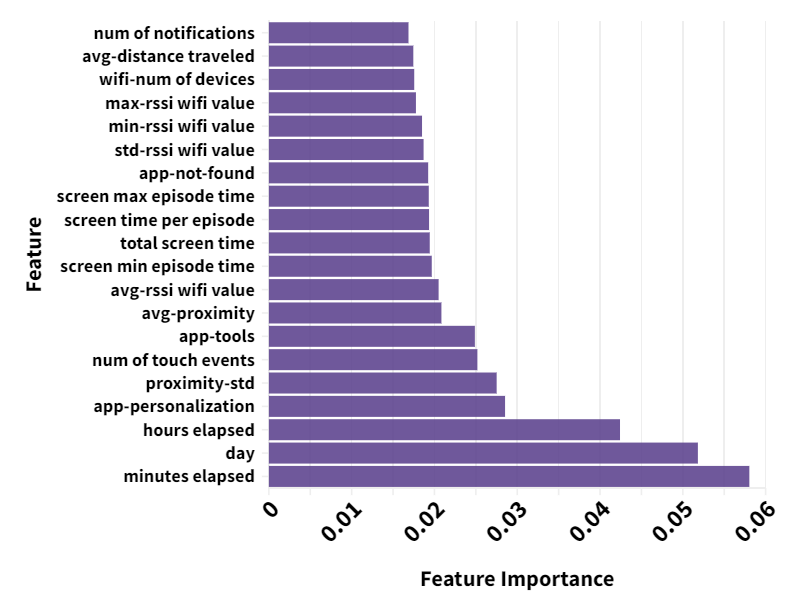}
        \caption{\wageesha{Gini Feature Importance values  of top 20 features generated using Random Forest for CBM of MUL dataset }}
        \label{fig:fi_clm_mul}
    \end{minipage}
\end{center}
\vspace{-0.2 in}
\end{figure*}

\lakmal{In this section, we explore the Gini feature importance for mood inference, generated using Random Forest classifiers for CBM with passive sensing features, using both MEX and MUL datasets. Figures~\ref{fig:fi_clm_mex} and \ref{fig:fi_clm_mul} plot the values of top features in the MEX and MUL datasets, respectively. We observed that certain features, such as the radius of gyration, battery level, and time of the day, have higher importance values in the CBM of the MEX dataset. Conversely, in the MUL dataset, features representing time of the day, like minutes elapsed and hours elapsed, exhibit higher importance. On one hand, the higher feature importance of the radius of gyration suggests an association with the user's mood close to eating episodes, or vice versa. For instance, some studies have noted that users may travel longer distances to find food, regardless of their geographical location \cite{ma2015food, kerr2012predictors}. Additionally, there is a notable correlation between mobility patterns, time of the day, and psychological constructs such as depression \cite{canzian2015trajectories}, which might explain this finding. The significance of the battery level in all model types in the MEX dataset, though not immediately clear, aligns with previous research suggesting its informativeness regarding eating and drinking events \cite{Bae2017, meegahapola2021one}. This could be attributed to the battery level providing contextual information about the user's location \cite{Bae2017, Meegahapola2021Survey}. Furthermore, time of the day has been proven useful in eating behavior-related tasks \cite{Biel2018, meegahapola2021one}, and the hour of the day is a reasonable feature to estimate mood, where negative moods tend to increase from morning to evening \cite{mislove2010pulse}. The accelerometer's feature importance values are considerably higher in the MEX dataset, suggesting that physical activity levels around eating episodes could inform moods. Previous studies have shown that mood can influence daily activity levels \cite{roshanaei2009longitudinal}, detectable using acceleration data \cite{rabbi2011passive}. Screen events in the MUL dataset also have moderate feature importance values. Sano et al. \cite{sano2013stress} demonstrated correlations between screen events and user stress levels using mobile phones. Past studies have shown that internet usage can provide valuable insights into users' mental well-being, such as depression screening \cite{yue2020automatic}. In the MUL dataset, WiFi-related features prominently rank among the top features. Another study indicated that increased internet usage can elevate users' depression and anxiety levels \cite{campbell2006internet}. Bradley et al. \cite{bradley2023stress} discussed the impact of apps like TikTok, Instagram, and YouTube on users' moods. They noted that students who spend more time on their smartphones tend to feel less happy, less relaxed, and more stressed, while those who frequently check their smartphones feel happier and more relaxed, regardless of the app used. These results reinforce the potential role of mobile phone sensor data in mood-while-eating inferences using only passive sensing data.}

In summary, these results of \textbf{RQ3} suggests that personalization is a key component in achieving better performance in mood-while-eating inference task. The proposed community-based technique helps to overcome challenges such as the lack of individual data and the cold-start problem.

\section{Discussion}\label{sec:discussion}

\subsection{{Summary of Results}}\label{subsubsec:discussion_summary_of_results}

In this study, we studied mood-while-eating inference using two datasets with passive smartphone sensing and self-report data. The summary of the main findings is as follows:

\textbf{RQ1: } \lakmal{The application of generic mood inference models may not be sufficiently accurate for mood inference across varied contexts. Specifically, when analyzing the model's performance on mood-while-eating reports, we observed that increasing their representation in the training set led to improved performance in the corresponding subset of the testing set. However, this enhancement was offset by a decrease in overall performance, indicating the model's difficulty in generalizing its predictions across different situated contexts. These results highlight the need to develop context-specific models for mood-while-eating inference to increase the accuracy of mobile food diaries and mobile health applications. Additionally, these findings draw attention to the potential sub-context shift, a factor that could impact many inferences related to mobile sensing and longitudinal behavior modeling.}

\textbf{RQ2: } \lakmal{The performance of mood-while-eating inference models was assessed using both population-level and hybrid modeling approaches. The population-level approach yielded an accuracy of 24.1\% for the MEX dataset and 82.4\% for the MUL dataset when using passive sensing data. Conversely, the hybrid modeling approach showed an accuracy of 54.6\% for the MEX dataset and 83.7\% for the MUL dataset. The results suggest that the hybrid modeling approach may be a more effective method for mood-while-eating inference. Additionally, we discussed the challenges in training fully personalized user-level models across both datasets, primarily due to insufficient data in certain classes.}

\textbf{RQ3: } We proposed a community-based personalization technique as a solution to the challenges faced by traditional mood inference models, such as a lack of individual data from certain classes and the cold-start problem. Through the implementation of this technique, we observed accuracies of 63.8\% (F1-Score: 62.5\%) and 88.3\% (F1-Score: 85.7\%) for the MEX and MUL datasets, respectively. These results demonstrate the potential of community-based personalization as an approach for addressing the limitations of conventional models in mood inference.

\subsection{{Implications}}\label{subsubsec:discussion_implecations}

\paragraph{Implications for Modeling} In this study, we presented an investigation of personalization as a technique to improve the performance of models in a three-class mood inference task that uses only smartphone sensing data to infer mood-while-eating. Our proposed personalization technique addresses the challenges of limited individual data in a dataset and the cold-start problem. Our results indicate that the use of personalization techniques is crucial for the improvement of performance in mood inference tasks. However, we acknowledge that the collection of individual data is a challenging task. Thus, future studies should focus on capturing more data from individual users to investigate further various personalization techniques that can improve the performance of mood inference models. Additionally, our study highlights the importance of context-specific models for mood inference. In this study, we focused on the specific context of eating episodes, and future research should explore other context-specific mood inference tasks. Furthermore, the application of domain adaptation techniques for multimodal sensor data has yet to be studied in depth \cite{xu2023globem, meegahapola2023generalization}. While domain adaptation has been extensively researched in the fields of computer vision, NLP, and speech, its application to multimodal sensor data is an area that warrants further investigation.

\paragraph{Implications for Applications} The results presented in this study provide valuable insights into the potential of utilizing passive sensing data to infer context-specific moods, specifically in the context of eating episodes. This has important implications for the design and development of mobile health applications and mobile food diaries, which could leverage this information to provide tailored interventions and feedback to users, in real time. For example, by identifying negative moods associated with eating episodes, mobile health applications could send notifications or suggestions to users who have a tendency to overeat in such states, thereby preventing unhealthy eating patterns. Additionally, the use of smartphone-based sensing data in this study highlights the potential for mobile application developers to create cost-effective solutions that do not require expensive wearable devices. As such, further research should be conducted to investigate other context-specific mood inference tasks, and to explore the potential of domain adaptation techniques in multimodal sensor data analysis. We also opened up another question: while capturing sensor data throughout the day and inferring generic mood could be useful, doing it with resource-limited devices raises some issues. Adding to the previous body of literature on inference mental well-being-related variables in specific contexts, we foresee that monitoring mood in certain contexts (i.e., mood-while-eating, mood-while-at-work, mood-while-driving) could have practical uses, be more energy efficient for phones, and overall add value to users.

\subsection{{Limitations and Future Work}}\label{subsubsec:discussionlimitations}

The field of mood inference using passive sensing has been an active area of research for over a decade \cite{Meegahapola2021Survey}. Our study, however, highlights the need for further experimentation in order to ensure that inferences work effectively in a diverse range of real-life contexts. In this particular case study, we focused on the context of eating occasions, yet there are numerous other contexts, such as drinking occasions and social gatherings, that could be explored~\cite{labhart2021ten}. Additionally, cultural diversity represents another important aspect to consider \cite{schelenz2021theory}. The datasets used in this study were collected in Mexico and eight other countries, and it is well-known that individual and food consumption behaviors can vary significantly across cultures \cite{kromhout1989food, de2015food, vereecken2006television}. Despite the second dataset being collected in eight countries, the limited number of labels in the negative class prevented us from building models for each country separately. Thus, it is imperative to investigate whether the inference and modeling techniques proposed in this study would generalize well to other countries, cultures, age groups, and contexts, such as eating, drinking, commuting, and being at home or at school. Conducting such experiments would greatly enhance the applicability of these models to real-world scenarios. \wageesha{Moreover, in this study, we explore the mood while eating. However, it is worth exploring the mood in other contexts as one study emphasized the importance of understanding the food, mood, and context \cite{morshed2022food}. Hence, future studies could focus on studying mood during other activities.}

\wageesha{Exploring how smartphones can provide feedback and interventions by considering a user's mood while eating brings its own set of challenges, including the potential risk of triggering individuals with eating disorders. Given the current state of mobile health research, the provision of interventions to users is still in a developmental phase. A previous study underscored the importance of appropriately timing the delivery of interventions, acknowledging that users might not always be receptive to processing these interventions \cite{mishra2021detecting}. Additionally, a study focusing on the National Eating Disorders Association's (NEDA) chatbot, which dispensed dieting advice, found that many users with eating disorders felt that the AI tool's advice exacerbated their conditions \cite{Wells_2023}. Future research should address these challenges and strive to conduct more comprehensive studies to yield conclusive insights.}

This study has several limitations. One such limitation is the reliance on cosine similarity as the similarity metric. While suitable for the datasets used in this study, other similarity measures like Mahalanobis distance \cite{lane2011enabling} may be more appropriate, depending on the dataset's nature. Additionally, the time window for sensor data aggregation around a self-report, which is one hour for the MEX dataset and 10 minutes for the MUL dataset, may not fully represent the users' mood for the entire duration.  It is important to note that this time window was chosen for aggregating sensor data to infer mood at the time of eating. Moreover, the study's inference pipeline, based on a well-established framework for eating behavior \cite{bisogni2007dimensions}, relies on mood reports captured through a single answer about valence, akin to previous studies in ubicomp \cite{likamwa2013moodscope}. This approach might have less validity compared to more comprehensive instruments used in clinical psychology for capturing affective states, like the PANAS \cite{crawford2004positive}. The single-question method was selected to minimize user burden, but future research could include more detailed instruments to study mood-while-eating at the episode level. Furthermore, this study focuses on inferring valence but not arousal \cite{russell1980circumplex}. Future directions could include studies inferring both arousal and valence at the time of eating. In addition, future research might explore more fine-grain mood inference tasks, such as five-class mood inferences, using larger datasets. 

Another limitation might arise from how we divided individual data for the HM and CBM model types. Kapoor and Narayanan \cite{kapoor2022leakage} highlighted the issue of temporal leakage in time series data, which is crucial to consider. Previous studies \cite{li2020extraction, liao2018just, mazankiewicz2020incremental} have addressed this by randomly splitting user data for training models. This splitting assumes no relationship between individual data points used in inference while ensuring no sensor data leaks between them. This assumption was explicitly stated in the original paper describing the MEX dataset \cite{meegahapola2021one}. Therefore, while we employed random user data splits, future research should approach such decisions cautiously, considering the nature of data, study designs, and underlying assumptions. In the results section, we observe a difference between the high accuracy/F1 scores and the lower AUC-ROC scores, which is likely due to class imbalance. The number of negative samples is significantly lower than the positive ones, affecting the AUC-ROC metric. Since AUC-ROC measures how well the model distinguishes between positive and negative labels, it is more sensitive to class imbalance, leading to lower scores. Finally,  another limitation of our study is the use of different test datasets (due to randomization) to evaluate the performance of the Population-Level Model (PLM), Hybrid Model (HM), and Community-Based Model (CBM). Ideally, for a fair comparison of the results across each model type, it would be best to use the same test dataset. Nonetheless, due to the inherent differences in the model types, it is challenging to assess their performance using a uniform test dataset. For example, while the PLM is developed using a leave-one-out strategy, the HM and CBM incorporate personalization strategies.

\section{Conclusion}\label{sec:conclusion}

In this paper, we aimed to investigate the relationship between mood and eating with mobile sensing data. We acknowledged the prior literature on mobile sensing that explored mood inference for generic moments and the need for more context-specific models. Additionally, we emphasized the significance of automatically inferring the mood-while-eating, as it could provide valuable feedback and interventions for mobile food diary users. Through examination of two datasets pertaining to the everyday eating behavior and mood of college students, we revealed that passive mobile sensing does not effectively infer the mood-while-eating using population-level modeling techniques. Furthermore, we demonstrated that user-level modeling techniques are limited in their applicability due to the scarcity of data. To address these limitations, we proposed a community-based personalization approach that allows for the training of partially personalized models even for target users with limited data. With this approach, we achieved an accuracy of 63.8\% with the Mexico dataset and 88.3\% with the multi-country dataset using passive sensing data. These results are overall promising on a three-class mood inference task. We also highlighted the importance of considering the application of mood inference in specific domains, such as mobile food diaries, where a comprehensive understanding of the user's mood is crucial to gaining a holistic understanding of eating behavior. We believe that inferring mood while eating could be a step towards making mobile food diaries more holistic and easier to adopt.

\begin{acks}

This work was funded by the European Union’s Horizon 2020 WeNet project, under grant agreement 823783. We acknowledge the use of ChatGPT and Grammarly for grammar correction. 

\end{acks}

\bibliographystyle{ACM-Reference-Format}
\bibliography{citations}

\newpage 
\appendix 
\section{APPENDIX}\label{sec:appendix}

\begin{figure*}[t]
\begin{center}
    \begin{minipage}[t]{0.49\textwidth}
        \centering
        \includegraphics[width=0.8\textwidth]{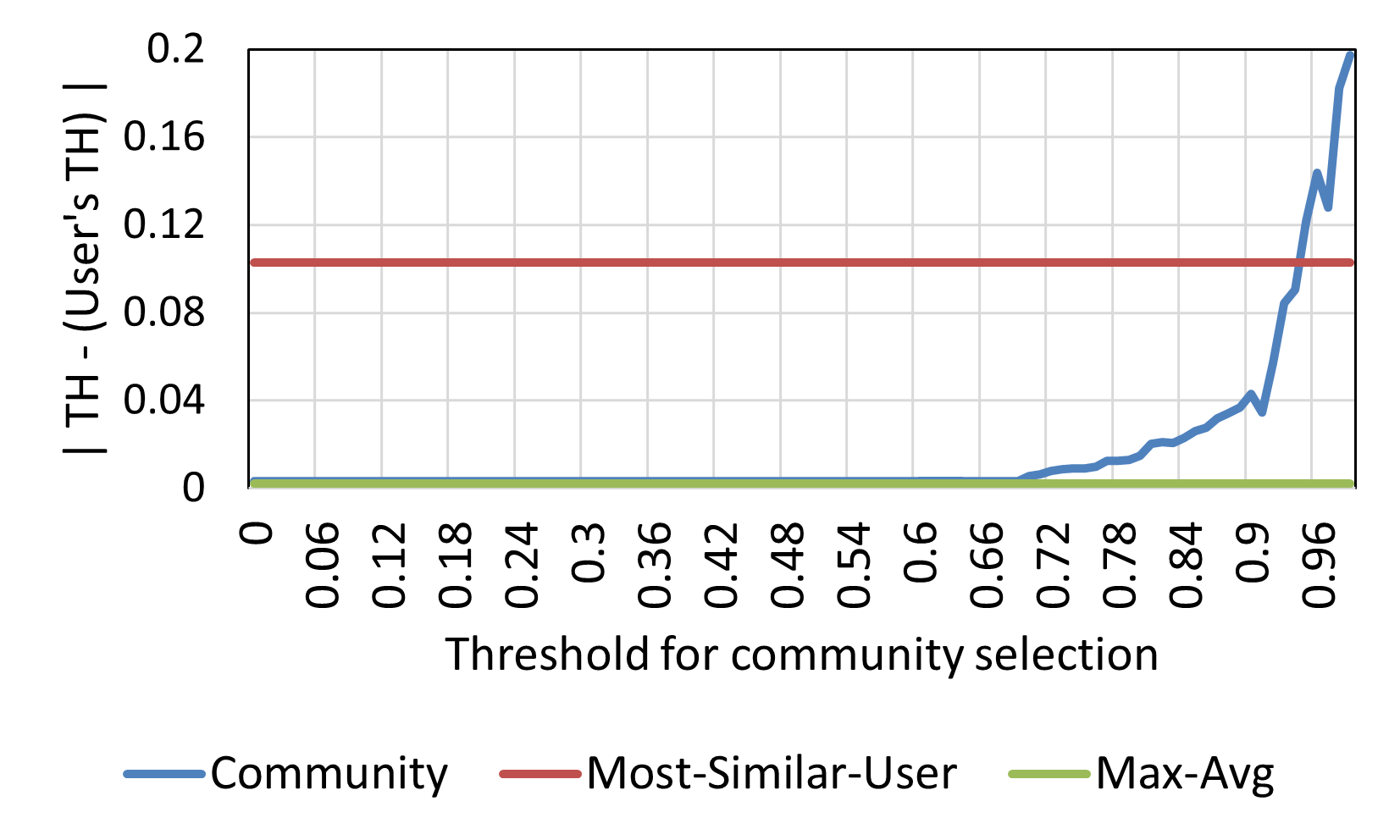}
        \caption{Averaged threshold distribution difference between target users and three threshold value distributions of MEX dataset}
        \label{fig:fi_threshold_selection}
    \end{minipage}
\end{center}
\vspace{-0.2 in}
\end{figure*}

\subsection{Selecting an Initial Threshold for Target Users using MEX dataset}
The threshold selection process when discovering the community is a bit time-consuming. For example, we used over twenty $th$ values and trained models and selected the $th$ that gave the highest accuracy. However, in real-world deployments, having a long list of thresholds in the very beginning could make the process of creating a model impractical. Hence, a way to estimate a good threshold is required. Given that each target user has different communities, corresponding to different thresholds (which can vary between 0 and 1), we need a threshold-searching algorithm to find the ideal threshold value. Figure~\ref{fig:fi_threshold_selection} shows the absolute value of threshold distribution difference between averaged optimum threshold value of target users (0.77) and three threshold value distributions: (i) Community: averaged threshold values of the community where each user in the community yields the highest accuracy. (ii) Most-Similar-User: averaged threshold values of the most similar user (obtained using the similarity metric) of each target user, where the most similar user yields the highest accuracy (this is a fixed value of 0.88). (iii) Max-Avg: Averaged threshold values corresponding to the maximum accuracy obtained for all the users (this is a fixed value of 0.78 -- see Figure~\ref{fig:clm_th_mex} MAX column). In each of the given three distributions, the one which shows the minimum difference (Max-Avg) can be taken as the first best option when selecting a threshold value for each new user, which can be used to optimize the threshold selection process. The conclusion from this analysis is, it is always better to use the Max-Avg of other's thresholds ($th$ = 0.78 for this set of users)  as a $th$ for any new user as it yields the lowest difference compared to the best threshold that can be obtained for any user. By using this average value as the threshold for any new user, the resource-consuming threshold selection process can be eliminated in the very beginning. In addition, this value provides a reasonably high performance compared to both PLM and HMs. However, note that this dynamic might change for other datasets and users.

\subsection{Results of the 5-Class Inference} \label{sec:appendix_5-class}

The results presented in Table~\ref{tab:5_class_inference_results_plm_hm} summarize the mean accuracy, F1-score, and AUC-ROC values for 5-class inference using Random Forest models on both PLM and HM. For the MEX dataset, PLM achieved an accuracy of 14.3\%, F1-score of 11.5, and AUC-ROC of 0.47. HM demonstrated comparatively higher performance with an accuracy of 46.4\%, F1-score of 46.7, and AUC-ROC of 0.51. The MUL dataset shows similar performance increases with the PLM and HM. The accuracy increased from 13\% to 62.6\% with PLM to HM model types. However, the performance of the 5-class models is notably lower compared to the corresponding 3-class models, indicating the increase in classification task complexity.

\begin{figure*}[t]
\begin{center}

    \begin{subfigure}[t]{0.49\textwidth}
        \centering
        \includegraphics[width=\textwidth]{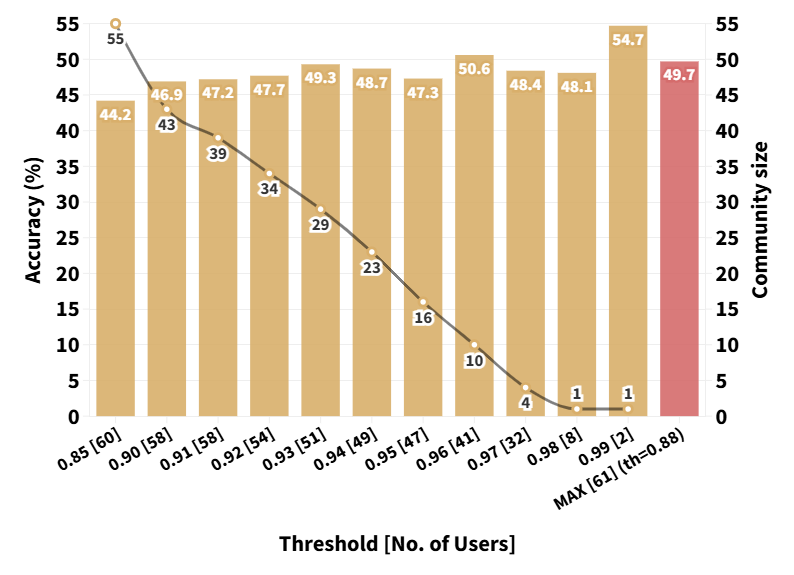}
        \caption{\wageesha{MEX dataset}}
    \end{subfigure}
    \begin{subfigure}[t]{0.49\textwidth}
        \centering
        \includegraphics[width=\textwidth]{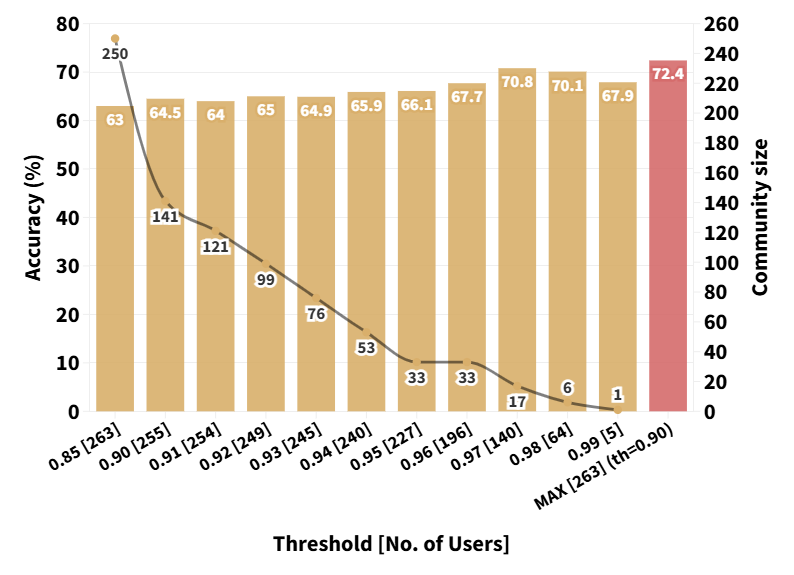}
        \caption{MUL dataset}
    \end{subfigure}
  
    \vspace{-0.1 in}
    \caption{Mean accuracy values(column values in the graphs) [with standard deviations] calculated using the random forest for CBM for multiple threshold values [No. of Users] and averaged community sizes (line values in the graphs)}
    \label{fig:5_class_clm_threshold_vs_acc}
\end{center}
\vspace{-0.2 in}
\end{figure*}

The CBM results with varying thresholds are illustrated in Figure~\ref{fig:5_class_clm_threshold_vs_acc}. For the MEX dataset, the highest accuracy of 54.7\% was achieved at a threshold of 0.99. In the rightmost column (MAX), the model's maximum accuracy was recorded as 49.7\%, with an average accuracy of 0.88 across all thresholds. For the MUL dataset, the maximum average accuracy (MAX) reached 72.4\%, which was obtained at a threshold value of 0.90. For the 5-class inference, the models performed reasonably well with the MUL dataset compared to the MEX dataset.

\begin{table}[t]
        \small
        \centering
        \caption{Mean (\={A}), Standard Deviation (A$_{\sigma}$) F1-Score (F1) and AUC-ROC value (AUC), of 5 class inference accuracies, calculated using random forest for the PLM and HM of mood inference task: \={A} (A$_{\sigma}$), F1, AUC}
        \resizebox{0.5\textwidth}{!}{%
        \begin{tabular}{l l l }

        \rowcolor{gray!10}
        \textbf{Feature Group}& 
        \textbf{PLM}&
        \textbf{HM} 
        \\

        
        \arrayrulecolor{black}
        \cmidrule{1-3}

        \textbf{(\# of Features)}& 
        \textbf{RF} &
        \textbf{RF}
        \\
        
        \textbf{}& 
        \textcolor{new}{\={A} (A$_{\sigma}$), F1, AUC}&
        \textcolor{new}{\={A} (A$_{\sigma}$), F1, AUC}
        \\

        \hline


        \arrayrulecolor{Gray}
\hline 
        
        MEX (40) & 
        14.3 (4.1), 11.5, 0.47 & 
        46.4 (15.6), 46.7, 0.51  
        \\
        

        \arrayrulecolor{Gray}
\hline 
        
        MUL (114) &
        13.0 (1.1), 9.5, 0.51 & 
        62.6 (8.1), 58.9, 0.32 
        \\

        \hline 
        
        \end{tabular}}
        \label{tab:5_class_inference_results_plm_hm}

\end{table}

\begin{algorithm}[t]
\vspace{6pt}
\hrule
\vspace{6pt}

\KwData{\\
$u$: a single user where $u \in U $ ; $U$ : all the users in the dataset.\\
$f$: a single feature where $f \in F $ ; $F$ : all the features in the dataset.\\
$D_{u}$: the dataset of a user $u$ where $D_{u} \in D$; $D$: complete dateset. \\
}
\KwResult{\\
$U_{aggr}$: user-based data aggregated matrix of size $(|U|\times|F|)$ \\
}
\vspace{\baselineskip} 

$U_{aggr} =  [\quad]_{|U|\times|F|} $ \tcp*{Initialize user-based data aggregated matrix}

\For{$u$ {\normalfont \textbf{in}} $U$}{ 
    $D_{u} = GetUserDataMatrix(D, u)$  \tcp*{Get all the row vectors(1x|F|) of the user u}
    $u_{aggr} = [\quad]_{1\times|F|}$ \tcp*{Initialize aggregated user vector}
    
    \For{$f$ {\normalfont \textbf{in}} $F$}{
        $u_{aggr}[f] = MeanValueOfColumn(D_{u}, f)$ \tcp*{Get mean of column f for each user}
    }

    $U_{aggr}[u] = u_{aggr}$
}
$Return(U_{aggr})$

\vspace{6pt}
\hrule
\vspace{6pt}
\caption{User-based Data Aggregation}
\label{alg:algorithm_user_based_data_aggregation}
\end{algorithm}


\begin{algorithm}[t]
\vspace{6pt}
\hrule
\vspace{6pt}

\KwData{\\
$U, u$ same as Algorithm~\ref{alg:algorithm_user_based_data_aggregation} \\
$T$: pre-defined threshold array. $th$: selected threshold value where $th \in T$\\
$U_{aggr}$: return value of Algorithm~\ref{alg:algorithm_user_based_data_aggregation} \\
$Sim_{U}$: similarity matrix of all the users of size (|U| $\times$ |U|) \\
$Sim_{u_{t}}$: similarity vector of the target user $u_{t}$ of size (1 $\times$ (|U|-1)) \\ 
}

\KwResult{\\
$U_{u_{t}}$: community of the target user $u_{t}$\\
}

\vspace{\baselineskip} 

$ Sim_{U} =  [\quad]_{|U|\times|U|}$ \tcp*{Initialize user-similarity matrix}
$ Sim_{U} =  CosineSimilarity(U_{aggr})$ \tcp*{Calculate the similarity metric among all user pairs}
$ Sim_{u_{t}} = GetUserSimilarityVector(Sim_{U}, u_{t})$ \tcp*{Get similarity vector of the target user}

$ U_{u_{t}} =  EmptyArray()$ \tcp*{Initialize the community matrix of the target user}

\For{$u$ {\normalfont \textbf{in}} $(U- u_{t})$}{
    $Corr_{u} = Sim_{u_{t}}[u]$ \tcp*{Get similarity value between $u_{t}$ and $u$}
    \If{$ Corr_{u} \geq  th $}{ 
        $ U_{u_{t}} = Append(U_{u_{t}},u)$ \tcp*{Append user $u$ to the community of the target user} 
    }
}

$Return(U_{u_{t}})$

\vspace{6pt}
\hrule
\vspace{6pt}
\caption{User-Level-Community Detection}
\label{alg:algorithm_user_community_detection}
\end{algorithm}


\begin{table*}
\centering
\caption{Summary of 40 Features Extracted from Smartphone Sensors.}
\label{table:allfeatures}
\resizebox{\textwidth}{!}{%
\begin{tabular}{>{\arraybackslash}m{3.8cm} >{\arraybackslash}m{16cm}}

\rowcolor{gray!20}
\textbf{\makecell[l]{Sensor (Acronym), \\ Number of features}}  & 
\textbf{Description: features} 
\\

\hline

\rowcolor{gray!8}
{Location, 1} &
considering movement within the time window \cite{Yue2014, Barlacchi2017} : radius\_of\_gyration
\\

{Accelerometer, 18} & 
features that represent the mean of all values and mean of absolute values (abs) were generated using accelerometer data for axes x, y, and z separately for the time window. time before (bef) and after (aft) the approx. eating time too were considered. considering the one-hour window: acc$_{x}$, acc$_{y}$, acc$_{z}$, acc$_{x}$abs, acc$_{y}$abs, acc$_{z}$abs; before and after T$_{anc}$: acc$_{x}$bef, acc$_{y}$bef, acc$_{z}$bef, acc$_{x}$abs\_bef, acc$_{y}$abs\_bef, acc$_{z}$abs\_bef, acc$_{x}$aft, acc$_{y}$aft, acc$_{z}$aft, acc$_{x}$abs\_aft, acc$_{y}$abs\_aft, acc$_{z}$abs\_aft
\\


\rowcolor{gray!8}
{Application, 10} &
app usage captures the behavior of participants \cite{likamwa2013moodscope, Santani2018}. the ten most frequently used apps in the dataset were selected. for each hour of consider, it was derived whether each app was used during the episode or not: facebook, whatsapp, instagram, youtube, chrome, spotify, android dialer, youtube music, google quick search box, microsoft launcher 
\\ 

{Battery, 6} &

battery and charging related features could capture the behavior of study participants \cite{Bae2017, Abdullah2016}. Therefore, the mean battery level was calculated. Furthermore, charging sources are also indicated (ac - alternative current, usb, or unknown): battery\_level, charging\_true, charging\_false, charging\_ac, charging\_usb, charging\_unknown
\\ 

\rowcolor{gray!8}
{Screen, 2} &
screen on and off events act as a proxy to user behavior \cite{Abdullah2016, Bae2017}: screen\_on, screen\_off
\\ 

{Date \& Time, 3} &

From the beginning of the day, the hour and the minute of the day was calculated \cite{Biel2018}. Whether the day is a weekend or not was derived to indicate behavioral differences for weekends and weekdays: hours\_elapsed, minutes\_elapsed, weekend 
\\

\hline    
    \end{tabular}
    }
\end{table*}


\begin{table*}
\centering
\caption{Summary of 114 Features Extracted from Smartphone Sensors}
\label{table:allfeatures_dataset2}
\resizebox{\textwidth}{!}{%
\begin{tabular}{>{\arraybackslash}m{3.8cm} >{\arraybackslash}m{16cm}}

\rowcolor{gray!20}
\textbf{\makecell[l]{Sensor, \\ Number of features}}  & 
\textbf{Description: features} 
\\

\rowcolor{gray!8}
{Location, 8} &
 considering movement within the time window: radius of gyration, distance traveled, mean altitude
\\

{Bluetooth, 10} & 
considering usage of Bluetooth feature: number of devices, mean/std/min/max
rssi (Received Signal Strength Indication – measures how close/distant other
devices are)
\\

\rowcolor{gray!8}
{WiFi, 6} &
considering usage of wifi: connected to a network indicator, number of devices, mean/std/min/max rssi
\\ 

{Cellular, 8} &
number of devices, mean/std/min/maxphone signal strength
\\ 

\rowcolor{gray!8}
{Notifications, 4} &
considering the notifications received to mobile phone: notifications posted, notifications removed
\\ 

{Proximity, 4} &
mean/std/min/max of proximity values
\\ 

\rowcolor{gray!8}
{Activity, 8} &
considering time spent doing activities: still, in\_vehicle, on\_bicycle, on\_foot, running, tilting, walking, other (derived using the Google activity recognition API)
\\ 

{Steps, 2} &
steps counter, steps detected
\\

\rowcolor{gray!8}
{Screen events, 6} &
considering users' screen interactions: number of episodes, mean/min/max/std episode time, total time
\\ 

{User presence, 1} &
considering time the user is present using the phone: user present time
\\

\rowcolor{gray!8}
{Touch events, 1} &
considering users' screen touch events
\\ 

{App events, 48} &
considering time spent on apps of each category derived from Google Play Store:action, adventure, arcade, art \& design, auto \& vehicles, beauty, board, books \& reference, business, card, casino, casual, comics, communication, dating, education, entertainment, finance, food \& drink, health \& fitness, house, lifestyle, maps \& navigation, medical, music, news \& magazine, parenting, personalization, photography, productivity, puzzle, racing, role playing, shopping, simulation, social, sports, strategy, tools, travel, trivia, video players \& editors, weather, word, not\_found
\\

\rowcolor{gray!8}
{Date \& Time, 5} &
considering hours elapsed and minutes elapsed from the beginning of the day and other time features such as day, week, month
\\

    \end{tabular}
    }
\end{table*}

\begin{table*}
\centering
\caption{Summary of participants of MUL dataset  (countries named in alphabetical order)}
\label{table:MUL_demographocs}
\resizebox{\textwidth}{!}{%
\begin{tabular}{ >{\arraybackslash}m{4cm} >{\arraybackslash}m{4cm} >{\arraybackslash}m{4cm} >{\arraybackslash}m{4cm} >{\arraybackslash}m{4cm}}

\rowcolor{gray!20}
\textbf{Country}  & 
\textbf{Participants}   & 
\textbf{Mean Age (std.)}   & 
\textbf{\% Women}   & 
\textbf{\# Self-reports} 
\\

\rowcolor{gray!8}
China &
41 &
26.2 (4.2) &
51 &
22,289
\\

Denmark & 
24 &
30.2 (6.3) &
58 &
10,010
\\

\rowcolor{gray!8}
India &
39 &
23.7 (3.2) &
53 &
4,233
\\

Italy &
240 &
24.1 (3.3) &
58 &
151,342
\\

\rowcolor{gray!8}
Mexico &
20 &
24.1 (5.3) &
55 &
11,662
\\

Mongolia &
214 &
22.0 (3.1) &
65 &
94,006
\\

\rowcolor{gray!8}
Paraguay &
28 &
25.3 (5.1) &
60 &
9,744
\\

UK &
72 &
26.6 (5.0) &
66 &
26,688
\\

\rowcolor{gray!20}
Total &
678 &
24.2 (4.2) &
58 &
329,974
\\

    \end{tabular}}
\end{table*}

\begin{figure}[H]
    \centering
    \includegraphics[width=0.5\textwidth]{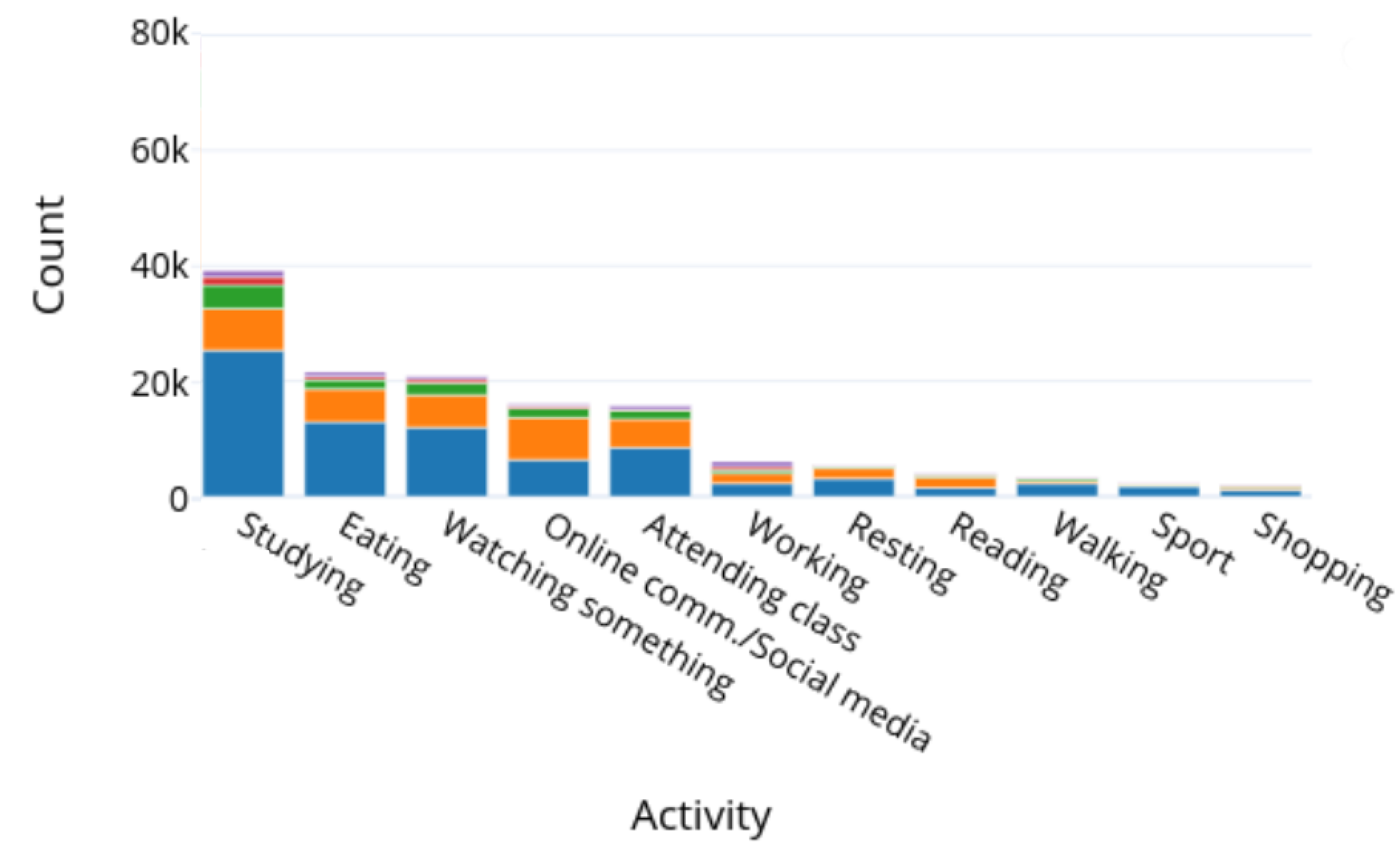}
    \caption{The distribution of the activities performed during daily life in the MUL dataset.}\label{smartphone_usage_daily_activities_MUL}
    \label{fig:fi_distribution_daily_activity_MUL}
\end{figure}

\end{document}